\documentclass[fleqn,usenatbib]{mnras}

\usepackage{newtxtext,newtxmath}
\usepackage[T1]{fontenc}
\DeclareRobustCommand{\VAN}[3]{#2}
\let\VANthebibliography\thebibliography
\def\thebibliography{\DeclareRobustCommand{\VAN}[3]{##3}\VANthebibliography}

\usepackage{graphicx}	
\usepackage{amsmath}	
\usepackage{xspace}
\usepackage{caption}
\usepackage{comment}

\newcommand{\angstrom}{\textup{\AA}\xspace}

\newcommand{\hd}{H$\delta$\xspace}
\newcommand{\hg}{H$\gamma$\xspace}

\newcommand{\niialma}{[N\,II]122$\mu$m\xspace}
\newcommand{\cii}{[C\,II]158$\mu$m\xspace}
\newcommand{\oiiialma}{[O\,III]88$\mu$m\xspace}

\newcommand{\lya}{Ly$\alpha$\xspace}
\newcommand{\hb}{H$\beta$\xspace}
\newcommand{\ha}{H$\alpha$\xspace}
\newcommand{\hz}{H$\zeta$\xspace}
\newcommand{\hep}{H$\epsilon$\xspace}
\newcommand{\het}{H$\eta$\xspace}

\newcommand{\heib}{HeI$\lambda5875$\xspace}
\newcommand{\heic}{HeI$\lambda3889$\xspace}

\newcommand{\oiilow}{[OII]$\lambda\lambda3726{,}3729$\xspace}
\newcommand{\oiilowa}{[OII]$\lambda3726$\xspace}
\newcommand{\oiilowb}{[OII]$\lambda3729$\xspace}
\newcommand{\oiiia}{[OIII]$\lambda4959$\xspace}
\newcommand{\oiiib}{[OIII]$\lambda5007$\xspace}
\newcommand{\oiiiab}{[OIII]$\lambda\lambda4959{,}5007$\xspace}
\newcommand{\oiiic}{[OIII]$\lambda4363$\xspace}
\newcommand{\neiiia}{[NeIII]$\lambda3869$\xspace}
\newcommand{\neiiib}{[NeIII]$\lambda3968$\xspace}
\newcommand{\neiiiab}{[NeIII]$\lambda\lambda3869{,}3968$\xspace}
\newcommand{\lmstar}{${\rm log}_{10}(M_{\star}/{\rm M}_{\odot})$}

\newcommand{\niiab}{[NII]$\lambda\lambda6548{,}6584$\xspace}

\newcommand{\siiab}{[SII]$\lambda\lambda6716{,}6731$\xspace}

\newcommand{\bfourteen}{B14-65666\xspace}
\newcommand{\pyneb}{\textlcsc{PyNeb}\xspace}

\newcommand\textlcsc[1]{\textsc{\MakeLowercase{#1}}}

\title[GA-NIFS: \bfourteen]{GA-NIFS: interstellar medium properties and tidal interactions in the evolved massive merging system B14-65666 at $\mathbf{z=7.152}$}

\author[G. C. Jones et al.]{Gareth C. Jones$^{1,2,3}$\thanks{E-mail: gj283@cam.ac.uk},
Rebecca A. A. Bowler$^{4}$,
Andrew J. Bunker$^{3}$,
Mirko Curti$^{5}$,
Santiago Arribas$^{6}$,\newauthor
Stefano Carniani$^{7}$,
Stephane Charlot$^{8}$,
Michele Perna$^{6}$,
Bruno Rodr\'iguez Del Pino$^{6}$,
Hannah \"{U}bler$^{9,1,2}$,\newauthor
Chris J. Willott$^{10}$,
Jacopo Chevallard$^{3}$,
Giovanni Cresci$^{11}$,
Eleonora Parlanti$^{7}$,
Jan Scholtz$^{1,2}$,\newauthor
Giacomo Venturi$^{7}$
\\
$^{1}$ Kavli Institute for Cosmology, University of Cambridge, Madingley Road, Cambridge CB3 0HA, UK\\
$^{2}$ Cavendish Laboratory, University of Cambridge, 19 JJ Thomson Avenue, Cambridge CB3 0HE, UK\\
$^{3}$ Department of Physics, University of Oxford, Denys Wilkinson Building, Keble Road, Oxford OX1 3RH, UK\\
$^{4}$ Jodrell Bank Centre for Astrophysics, Department of Physics and Astronomy, School of Natural Sciences, The University of Manchester, Manchester, M13 9PL, UK\\
$^{5}$ European Southern Observatory, Karl-Schwarzschild-Strasse 2, 85748 Garching, Germany\\
$^{6}$ Centro de Astrobiolog\'{i}a (CAB), CSIC-INTA, Ctra. de Ajalvir km 4, Torrej\'on de Ardoz, E-28850, Madrid, Spain\\
$^{7}$ Scuola Normale Superiore, Piazza dei Cavalieri 7, I-56126 Pisa, Italy\\
$^{8}$ Sorbonne Universit\'{e}, CNRS, UMR 7095, Institut d'Astrophysique de Paris, 98 bis bd Arago, 75014 Paris, France\\
$^{9}$ Max-Planck-Institut f\"ur extraterrestrische Physik (MPE), Gie{\ss}enbachstra{\ss}e 1, 85748 Garching, Germany\\
$^{10}$ NRC Herzberg, 5071 West Saanich Rd, Victoria, BC V9E 2E7, Canada\\
$^{11}$ INAF - Osservatorio Astrofisco di Arcetri, largo E. Fermi 5, 50127 Firenze, Italy\\
}

\date{Accepted 2026 February 15. Received 2026 February 13; in original form 2024 December 19.}
\pubyear{\the\year{}}

\begin{document}
\label{firstpage}
\pagerange{\pageref{firstpage}--\pageref{lastpage}}
\maketitle

\begin{abstract}
We present JWST/NIRSpec IFU observations of the $z=7.152$ galaxy system \bfourteen, as part of the GA-NIFS survey. Line and continuum emission in this massive system ($\log_{10}(M_*/M_{\odot})=9.8\pm0.2$) is resolved into two strong cores surrounded by diffuse emission, as seen in recent JWST/NIRCam imaging. Our dataset contains detections of \oiilow, \neiiiab, Balmer lines, \oiiiab, \heib, and weak \oiiic. Each spectrum is fit with a model that consistently incorporates interstellar medium conditions (i.e., electron temperature, $T_{\rm e}$, electron density, $n_{\rm e}$, and colour excess, $E(B-V)$). The resulting line fluxes are used to constrain the gas-phase metallicity ($Z_{\rm g}\sim0.2-0.3$ solar) and \hb-based SFR for each region. Common line ratio diagrams (O32-R23, R3-R2, Ne3O2-R23) reveal that each line-emitting region lies at the intersection of low- and high-redshift galaxies, suggesting low ionisation and higher metallicity compared to the predominantly lower-mass galaxies studied with the JWST/NIRSpec IFU so far at $z>5.5$. Spaxel-by-spaxel fits reveal evidence for both narrow (FWHM~$<400$\,km\,s$^{-1}$) and broad (FWHM~$>500$\,km\,s$^{-1}$) line emission, the latter of which likely represents tidal interaction or outflows. Comparison to ALMA \cii and \oiiialma data shows a similar velocity structure, and we explore optical-far infrared diagnostics. The two core galaxies both lie on the mass-metallicity relation at $z>4$, but show contrasting properties (e.g., $M_*$, $Z_{\rm g}$), suggesting distinct evolutionary pathways. Combining the NIRSpec IFU and ALMA datasets, our analysis opens new windows into the merging system \bfourteen.
\end{abstract}

\begin{keywords}
galaxies: high-redshift -- galaxies: interactions -- galaxies: ISM -- galaxies: kinematics and dynamics
\end{keywords}



\section{Introduction}

The first 1\,Gyr of the Universe ($z>6$) was a unique era of cosmic time. The first galaxies emerged (e.g., \citealt{lapo21,curtis23,carn25}) and started the process of ionising their surroundings (e.g., \citealt{naka23,bana18}). Galaxy formation was rapid and fruitful, with high galaxy merger rates (e.g., \citealt{duan25}), and evidence for increasing volume density of molecular gas (e.g., \citealt{riec19,arav24,rago24}) and star formation rate (SFR; e.g., \citealt{grup20,khus21,trai24}) with increasing cosmic time, based on extrapolation from $3<z<6$. 

The first $z>6$ galaxies were detected less than three decades ago (e.g., \citealt{fan01,hu02,stan03,tani05}), but through the advent of cutting edge telescopes, the redshift frontier has already been extended to within the first $\sim300$\,Myr of the Universe ($z>13$; e.g., \citealt{carn24a,koko24,wits24}). Morpho-kinematic studies of high-redshift galaxies with ALMA revealed a population of relaxed rotating disk galaxies (e.g., \citealt{neel20,lell21,herr22,rowl24}), but many observations revealed clumpy or merging systems (e.g., \citealt{carn18,jone21,lee25}). This was supported by the results of zoom-in cosmological simulations, which showed that bright $z>6$ galaxies accreted mass through frequent interactions with lower-mass satellites and/or collisions (e.g., \citealt{koha19,pall22}). However, most spectroscopic ALMA observations featured low spatial resolution ($\sim1''$, corresponding to $\sim6$\,kpc at $z=6$).

This picture was made even clearer with the advent of JWST \citep{gard23} and the integral field unit (IFU) on the NIRSpec instrument (\citealt{boke22,jako22}). With a narrow point spread function ($\sim0.1-0.2''$ or $0.6-1.2$\,kpc at $z=6$; \citealt{deug23a}), the NIRSpec IFU is able to map multiple bright rest-UV and rest-optical lines at $z>4$ (e.g., \ha, \oiiiab). While JWST follow-up observations of some galaxies classified as rotators based on previous rest-frame far-infrared (FIR) observations confirmed their nature (e.g.; ALESS73.1, \citealt{lell21,parl24aless}; GN20, \citealt{hodg12,uble24}), other galaxies were found to feature clumpy morphologies, close companions, and sometimes outflows (e.g.; HZ10, \citealt{jone17,jone24hz10}; HZ4, \citealt{herr22,parl24hz}; COS3018, \citealt{smit18,scho24}). Due to the detection of multiple emission lines, these data were used to map morphology, kinematics, and the interstellar medium (ISM) conditions (e.g., electron density $n_{\rm e}$ and temperature $T_{\rm e}$, and gas-phase metallicity $Z_{\rm g}$) of each target.

A number of JWST/NIRSpec IFU observations have already detailed the properties of $z>6$ galaxies (e.g., \citealt{hash23,deca24,fuji24,liu24,lyu24,mess24,scho24,vent24}). Many of these have been acquired as part of the Galaxy Assembly with NIRSpec Integral Field Spectroscopy (GA-NIFS)\footnote{\url{https://ga-nifs.github.io/}} Guaranteed Time Observation (GTO) survey (Cycles 1 and 3; PIs R. Maiolino \& S. Arribas; e.g. \citealt{mars23,mars24,arri24,jone24,marc24a,marc24b,scho24,uble24zs7}). This survey, which includes 55 targets between $2<z<11$, was designed to exercise the power of the NIRSpec IFU through observations of a variety of well-studied galaxies (e.g., star-forming galaxies, active galactic nucleus [AGN] host galaxies, quasars). In this work, we present the GA-NIFS observations of the $z\sim7.152$ Lyman break galaxy (LBG) \bfourteen.

A key factor in the discovery of \bfourteen was the Ultra Deep Survey with the VISTA (Visible and Infrared Survey Telescope for Astronomy; \citealt{emer10}) telescope (UltraVISTA; \citealt{mccr12}), which targeted the COSMOS field \citep{scov12} in $Y$, $J$, $H$, and $K_s$ near-infrared bands. Combining UltraVISTA DR1 with other optical and infrared data, \citet{bowl12} first reported \bfourteen as one of only four robust $z>6.5$ galaxies from a sample of $>2\times10^5$\,detected sources. \citet{bowl14} performed a follow-up photometric SED modelling analysis combining UltraVISTA DR2 and additional data, improving the accuracy of the photometric redshift of this source.

The first resolved image of this source was taken with HST/WFC3 F140W \citep{bowl17}, where two closely separated clumps (aligned nearly east-west) are visible. 
ALMA follow-up observations of the source were first performed in~\citet{bowl18}, where \bfourteen was the only dust continuum detected source in Band 6 in a sample of six luminous LBGs.
\citet{hash19} confirmed the dust continuum detection and morphology of this source. In addition, they detected the spatially resolved \cii ($L_{\rm [CII]158\mu m}=(11.0\pm1.4)\times10^8\,L_{\odot}$) and \oiiialma ($L_{\rm [OIII]88\mu m}=(34.4\pm4.1)\times10^8\,L_{\odot}$) FIR lines, which provided spectroscopic redshift confirmation ($z_{\rm sys}=7.1520\pm0.0003$)\footnote{Because it was the first galaxy in the Epoch of Reionisation (EoR) with spatially resolved emission in three powerful tracers (\cii, \oiiialma, and dust continuum emission), it was given the alternative alias of `Big Three Dragons' after a combination of tiles in the game \textit{Mahjong}.}. Modelling of the two-point dust SED yielded $T_{\rm dust}\sim50-60$\,K and $L_{\rm FIR}\sim10^{12}$\,L$_{\odot}$. The dust continuum emission appears to be elongated, potentially aligned with the two components (see also~\citealp{bowl22} who find the dust to be between the two UV clumps).  Based on these properties, Hashimoto et al. suggest that \bfourteen is a merger-driven starburst.

Follow-up ALMA observations did not return a significant \niialma emission ($<8.1\times10^7\,L_{\odot}$), but strong underlying FIR continuum emission \citep{suga21}. Detailed dust SED modelling returned similar $T_{\rm dust}$ and $L_{\rm FIR}$ constraints as \citet{hash19}, with an additional estimate on the dust mass of $M_{\rm dust}\sim10^7\,M_{\odot}$. Similarly, ALMA observations of reliable molecular gas tracers (CO(7-6), CO(6-5), [CI](2-1)) and their underlying dust continuum emission resulted in non-detections \citep{hash23}. The molecular gas mass estimates implied by these non-detections are in agreement with the mass found through scaling relations based on the detection of \cii \citep{madd20}, FIR continuum emission \citep{li19}, and the kinematics of the \cii emission \citep{hash19}: $M_{\rm mol}=10^{8.7-11.0}\,M_{\odot}$. 

JWST/NIRCam imaging of \bfourteen was presented by \citet{suga24} as part of the `Reionization and the ISM/Stellar Origins with JWST and ALMA' (RIOJA) project (JWST GO1 PID1840; PIs: J. \'{A}lvarez-M\'{a}rquez and T. Hashimoto; \citealt{hash23b}). 
From the flux excess between the F356W and F444W bands they infer the presence of strong \oiiib emission (rest-frame equivalent widths of $1000$--$2000$\,\AA; as expected from the observed Spitzer colours;~\citealp{bowl14}).
Interestingly, the inferred \oiiib emission peaks in the Eastern component of the source, where the rest-frame UV emission is weaker.
Through SED fitting of the resolved NIRCam and ALMA multi-band fluxes, \citet{suga24} derive a stellar mass of \lmstar$ = 9.78^{+0.16}_{-0.18}$ and a relatively low metallicity of $(Z_{*}/Z_{\odot}) = 0.21^{+0.05}_{-0.04}$.
The extremely high spatial resolution of the NIRCam data reveals that the Eastern component is compact, while the Western component is spatially extended.  
Additional low surface brightness features are identified as tidal tails. Recent JWST/MIRI MRS observations \citep{prie25} revealed strong \ha emission from the two core components, with component E exhibiting a higher dust attenuation.

In this work, we exploit NIRSpec IFU observations from $\lambda_{\rm obs}=2.9-5.3$\,$\mu$m to directly map the strength of the rest-frame optical emission lines across the source. In Section \ref{datdes}, we describe both our new JWST/NIRSpec IFU data for \bfourteen as well as archival data explored in this work. The NIRSpec data are then analysed in Section \ref{NSanalysis}. We interpret and discuss these results in Section \ref{discsec}  and conclude in Section \ref{concs}.
We use a standard concordance cosmology (i.e., $H_0=70$\,km\,s$^{-1}$\,Mpc$^{-1}$, $\Omega_m=0.3$, $\Omega_{\Lambda}=0.7$) throughout, where $0.1''$ corresponds to $\sim$0.52\,kpc at $z=7.152$. A \citet{chab03} initial mass function (IMF) with an upper mass limit of 100\,$M_{\odot}$ is assumed.

\section{Data description}\label{datdes}

\subsection{JWST/NIRSpec IFU}\label{ifu_data}
The data analysed here originated from GA-NIFS observations as part of project 1217 (PI: N. Luetzgendorf; details in Table \ref{JWST_spec}). An eight-point `MEDIUM' dither pattern was used, with a starting point of `1'. Data were taken at medium spectral resolution (G395M/F290LP; $R\sim1000$), covering $\lambda_{obs}=2.871-5.270\,\mu$m ($\lambda_{rest}\sim3522-6465\angstrom$ at $z=7.1520$). No \textit{leakcal} or background exposures were taken. The data were calibrated with the STScI pipeline (v1.15.0, CRDS 1241), with custom outlier rejection (\citealt{deug23a}), custom masks for contamination by cosmic ray strikes (`snowballs')
and open MSA shutters leakage, 1/f noise corrections for count-rate maps, and drizzle weighting to create a data cube with spatial pixels (spaxels) of width $0.05''$ (see \citealt{pern23a} for full details of reduction). To ease spatial comparison of emission at different wavelengths, we homogenise the PSF of the cube (see Appendix \ref{psfapp}).

\begin{table}
\caption{JWST NIRSpec/IFU observation properties.}
\centering
\begin{tabular}{c|c}
Grating/Filter & G395M/290LP\\ \hline
Readout Pattern & NRSIRS2 \\
Groups/Int & 25 \\
Ints/Exp & 1\\
Exposures/Dithers & 8\\
Total Time [ks] & 14.7\\
\end{tabular}
\label{JWST_spec}
\end{table}

Previous analyses of JWST/NIRSpec IFU data (e.g., \citealt{jone24,parl24hz,uble24}) found astrometric errors due to the pointing accuracy of JWST ($\sim0.1''$, \citealt{rigb23}). To correct this, we align our data to the \textit{Gaia} DR3 reference frame (\citealt{gaia16,gaia21}) through a comparison to JWST/NIRCam data (see Appendix \ref{astrom_app}).

No background exposures were included in these observations, so we manually performed a background subtraction. For each wavelength of our data cube, we first temporarily mask the containing significant emission, and then use the \textlcsc{photutils} task \textit{Background2D} with a box size of 10\,px and filter size of 5\,px to create a spatially smoothed background map. These background maps are merged into a background cube, which is spectrally smoothed using a median filter of width 25\,channels. We verify that this background cube does not contain any spectral features, and subtract it from the observed data cube.

JWST/NIRSpec IFU observations of high-redshift galaxies revealed the presence of sinusoidal artefacts (`wiggles') in single-spaxel spectra (e.g., \citealt{pern23a,deca24,uliv24}). These wiggles, which were caused by under-sampling of the PSF, were more noticeable for drizzled maps featuring bright point sources coincident with extended emission (i.e., AGN and QSOs). We have followed the procedure of \citet{pern23a} to inspect our data for these wiggles, finding little evidence of strong wiggles in our data. While their presence may be detected as high-frequency residuals in spectral fits (see analysis in Section \ref{intspec}), they are very low-level ($\lesssim2\%$, or $<2\sigma$), so a correction is not applied.

\subsection{ALMA}

Due to the bright nature of this high-redshift source, ALMA has been used to target its emission in CO(6-5), CO(7-6), and [CI](2-1) (2018.1.01673.S, PI Hashimoto), \cii (2016.1.00954.S, PI Inoue), \oiiialma (2016.1.00954.S and 2017.1.00190.S, PI Inoue), and \niialma (2019.1.01491.S, PI Inoue). In order to compare the distribution of these lines (with their underlying continuum emission) and our data, we re-image each dataset. 

To begin, we download all data from the ALMA data archive and run the calibration pipeline provided by NRAO staff\footnote{2019.1.01491.S was calibrated directly by the ALMA helpdesk.}. The calibrated data for \bfourteen was split out, and a 30\,s time average was performed to ease data usage. Data were combined for each band (3, \citealt{clau08,kerr14}; 6, \citealt{kerr04,edis04}; 7, \citealt{mahi12}; 8, \citealt{seki08}). 

Continuum images were created with the CASA task \textit{tclean} in multi-frequency synthesis (MFS) mode, natural weighting, cell sizes of $1/5^{\rm th}$ the FWHM of the minor axis of the synthesised beam, and excluding all data within $\pm1000$\,km\,s$^{-1}$ of expected line emission. `Clean' continuum maps were created by cleaning down to $3\times$ the RMS noise level of initial `dirty' images.

No significant line emission is detected for CO(6-5), CO(7-6), [CI](2-1), or \niialma (as originally found by \citealt{hash23} and \citealt{suga21}), while \cii and \oiiialma are strongly detected. For the latter two lines, we create clean line+continuum cubes using CASA \textit{tclean} in `cube' mode with natural weighting and cell sizes of $1/5^{\rm th}$ the FWHM of the minor axis of the synthesised beam. Each cube is continuum subtracted in the image plane\footnote{Due to the resolved, relatively weak nature of the line emission, this is preferable to visibility-space continuum subtraction.} using the CASA task \textit{imcontsub}. The continuum level in each spaxel is found by fitting the line-free channels in the spectrum with a first-order polynomial (using the same $\pm1000$\,km\,s$^{-1}$ range as above), and the resulting continuum emission is removed. This results in a continuum-free data cube.

\subsection{JWST/NIRCam}\label{NCsec}

\bfourteen was observed with JWST/NIRCam in multiple filters as part of Project 1840 (PI J. Alvarez-Marquez ; see detailed analysis of \citealt{suga24}). Of these, we use only the F356W filter for astrometric correction (see Appendix \ref{astrom_app}). This NIRCam image was downloaded from the Mikulski Archive for Space Telescopes (MAST\footnote{\url{https://mast.stsci.edu/portal/Mashup/Clients/Mast/Portal.html}}). By inspecting the position of a \textit{Gaia} DR3 reference star in the F356W image, we determine a spatial offset and align it to \textit{Gaia} frame. 

\section{NIRSpec analysis}\label{NSanalysis}

Our JWST/NIRSpec IFU data cube allows us to explore line and continuum emission from \bfourteen on a spatially and spectrally resolved basis. In this Section, we first characterise the field by fitting integrated spectra using emission-based apertures (Section \ref{intspec}). We then exercise the power of the NIRSpec IFU by extracting and fitting spectra from each individual spaxel (Section \ref{spaxelbyspaxel}), opening a more detailed view into the ISM conditions and morpho-kinematics of \bfourteen.

\subsection{Spatially integrated spectral analysis}\label{intspec}
First, we characterise the emission in the field by extracting integrated spectra from our NIRSpec data cube using the apertures created by the recent JWST/MIRI analysis of \bfourteen by \citet{prie25}, which are listed in Table \ref{aper_table}. These apertures were created to characterise the two bright cores of emission (named `E' and `W'), as well as their sum (`E+W') and a larger aperture that is designed to capture all significant line emission (`Total'). We proceed with apertures `E' and `W' in this subsection. Throughout this work, we will refer to these apertures as `cores'. After extracting each spectrum, we perform an aperture loss correction (see Appendix \ref{psfapp} for details).

\begin{table}
\centering
\begin{tabular}{c|cccc}
Aperture & RA           & Dec          &  HWHM              & PA\\ 
Name     & [$^{\circ}$] & [$^{\circ}$] & [$''\times''$]     & [$^{\circ}$]\\\hline
E        & 150.4195767  & 1.9146156    & $0.211\times0.201$ & 112.83\\
W        & 150.4194613  & 1.9145781    & $0.228\times0.177$ & 112.83\\
E+W      & 150.4195242  & 1.9145981    & $0.456\times0.343$ & 85.83\\
Total    & 150.4195242  & 1.9145981    & $0.9\times0.9$     & -\\ \hline
\end{tabular}
\caption{Details of elliptical apertures used in this work, as adopted from JWST/MIRI analysis of \citet{prie25}. We include the central position (RA and Dec), half width at half maximum (HWHM) along the major and minor axes, and position angle (PA). We primarily utilise the first two apertures, which are designed to capture emission from the two core galaxies.}
\label{aper_table}
\end{table}

\subsubsection{Model description}\label{moddes}

\begin{figure*}
\centering
\includegraphics[width=\textwidth]{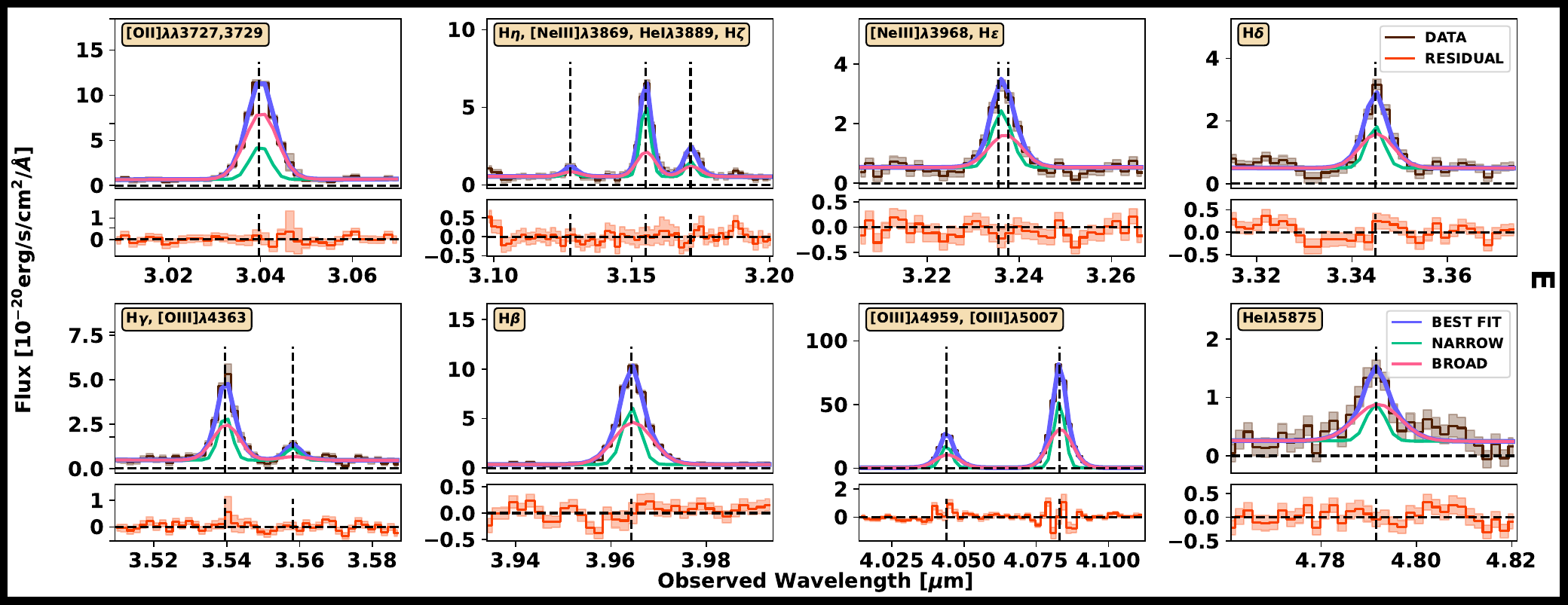}
\includegraphics[width=\textwidth]{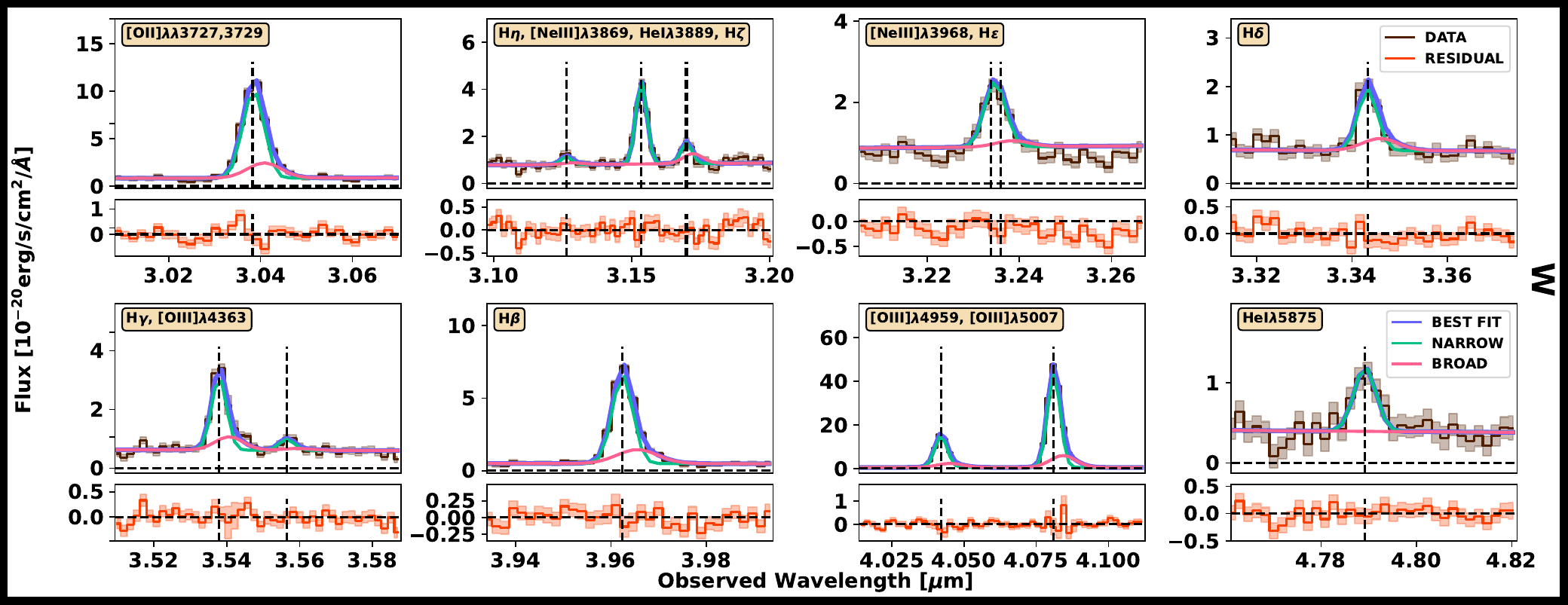}
\caption{Spectra extracted from our JWST/NIRSpec IFU data cube using the two primary apertures (E and W, see Table \ref{aper_table}). Aperture loss corrections (see Appendix \ref{psfapp}) have been applied to each spectrum. We zoom in around each emission line and present the observed spectrum, the best-fit model, and its narrow and broad components. The residual is included in the lower panel. The redshifted wavelengths of each fit line (using the best-fit redshift of the narrow component) are marked with vertical dashed lines.} 
\label{spec_1}
\end{figure*}

For the spectrum from each aperture, we assume that the continuum is described by a single power law model. Since the bluest edge of our data ($\lambda_{\rm rest}\sim 3620\angstrom$) extends slightly bluewards of the Balmer break ($\lambda_{\rm rest}\sim 3646\angstrom$), it is possible that the continuum is affected by Balmer absorption features and/or in a change in slope bluewards of \neiiia, as seen in quenched galaxies (e.g., \citealt{loos24,deug25}). However, we do not detect such features in our data, and the simple model returns low residuals (see Figure \ref{spec_1}).

To account for line emission, we include contributions from \oiilow, \neiiiab, Balmer lines (\het, \hz, \hd, \hep, \hb), \heic, \oiiic, \oiiiab, and \heib in our model. Because some lines exhibit asymmetric wings, we follow other investigations of high-$z$ galaxies with NIRSpec and fit each emission line with two components: one narrow ($\rm FWHM\le350$\,km\,s$^{-1}$) and one broad, representing tidal features or an outflow ($\rm 350\le FWHM\le1000$\,km\,s$^{-1}$; e.g., \citealt{lamp24,rodr24}). A one-Gaussian fit returns large residuals for each of the strong lines (\oiiib, \oiiia, \oiilow), suggesting that a two-Gaussian fit is needed. This approach is preferred to Voigt profiles (e.g., \citealt{jone24}), as it allows for the fitting of asymmetric lines. 

Due to the spectral blending of multiple lines (e.g., \heic and \hz, \neiiib and \het), a global fit of each extracted spectrum results in poor constraints on line fluxes. To lower the degrees of freedom in our model, we make the standard assumption of case B recombination (\citealt{bake38}; e.g., \citealt{hu24,soli24,vent24})\footnote{Although we note that the applicability of this assumption for high-redshift galaxies has been called into question (e.g., \citealt{scar24}).} and use the python package \pyneb \citep{luri15} to calculate line ratios based on ISM conditions (i.e., electron temperature $T_{\rm e}$, electron density $n_{\rm e}$, and B-V colour excess E(B-V) ; see further discussion in Appendix \ref{pnapp}). 

We fit each extracted spectrum in stages. First, we examine a broad wavelength range that includes only the strongest line (\oiiib) and the underlying continuum. Using \textlcsc{lmfit}, we fit for the continuum properties (i.e., power law slope and normalization) and the fluxes ($F_{\rm [OIII]\lambda5007,N}$, $F_{\rm [OIII]\lambda5007,B}$), widths ($FWHM_{\rm N}$, $FWHM_{\rm B}$), and redshifts ($z_{\rm N}$, $z_{\rm B}$) of the narrow and broad components of \oiiib. An initial fit is performed where all variables are allowed to vary. The results of this fit are inspected, and components with poorly determined (i.e., $<3\sigma$) or negligible fluxes (i.e., $<10^{-21}$\,erg\,s$^{-1}$\,cm$^{-2}$) are removed from the model before the fit is repeated. This process repeats until \textlcsc{lmfit} reports convergence. In the following steps, we assume that these linewidths and redshifts are applicable to each line, and apply the component rejection criteria at each step. 

Next, we isolate the wavelength ranges around the strongest observed Balmer line (\hb) and helium line (\heib), and find the fluxes of the narrow and broad components of each. While the ratio of the two lines in the \oiilow doublet is a useful tracer of electron density (e.g., \citealt{kewl19}), their close separation requires high spectral resolution for detailed fitting ($R\gtrsim2700$; \citealt{comp13}). This is achievable with high-resolution JWST/NIRSpec gratings (e.g., \citealt{rodr24,chis24}), but our resolution is a factor of $\sim3$ more coarse. Therefore, we fix $n_{\rm e}$ to the fiducial value of $n_{\rm e}=200$\,cm$^{-3}$ \citep{suga24}, calculate the ratio of \oiilowa/\oiilowb using \pyneb, and fit for the fluxes of each line with \textlcsc{lmfit}.

For the next step, we consider a range that contains the next two strongest observed Balmer lines (\hg and \hd), as well as two oxygen lines (\oiiia, \oiiic). The flux ratios of each Balmer line pair are related through ISM conditions and dust extinction (here parametrised as E(B-V), where we assume a \citealt{calz00} dust attenuation curve), while \oiiic/\oiiib is a strong function of $T_{\rm e}$, and \oiiia/\oiiib (which is fixed by atomic physics to be 2.984) is only a function of $E(B-V)$. Thus, by taking our previously measured \hb flux and assuming a fiducial density, we are able to fit for the fluxes of the narrow and broad components of these four lines self-consistently with few parameters ($T_{\rm e,N}$, $T_{\rm e,B}$, $E(B-V)_{\rm N}$, $E(B-V)_{\rm B}$). The work of \citet{suga24} estimated the dust reddening of the stellar continuum in this system to be $A_{\rm V}^*\sim0.3-1.2$, which corresponds to ionised gas colour excess of $E(B-V)\sim0.2-0.7$. Thus, if our fits return large uncertainties on E(B-V) or $T_{\rm e}$ (i.e., best fit values that are $<3\times$ the reported uncertainty), then we assume a value of $E(B-V)=0.2$ (i.e., the smallest value found by \citealt{suga24}) or $T_{\rm e}=1.2\times10^4$\,K \citep{suga24}, respectively.

As a final step, we fit for the flux of \neiiia. The flux of \neiiib is calculated using the ratio of \neiiib/\neiiia (as fixed by atomic physics; 0.301), and we use the previously determined ISM properties and line fluxes to calculate the fluxes of each remaining Balmer line (\het, \hep) and helium line (\heic).

Our fits take the fiducial line spread function (LSF) of NIRSpec\footnote{As recorded in the JWST documentation; \url{https://jwst-docs.stsci.edu/jwst-near-infrared-spectrograph/nirspec-instrumentation/nirspec-dispersers-and-filters}} into account. This varies from $\sim400$\,km\,s$^{-1}$ (FWHM) at the wavelength of \oiilow to $\sim300$\,km\,s$^{-1}$ at the wavelength of \heib. We experimented with allowing the LSF to vary from the fiducial value by $50\%$ (e.g., \citealt{degr24}), but the resulting fits were not well-constrained. The LSF-corrected FWHM and redshift of all lines are fixed for each narrow and broad component. 

For each spectrum, our model therefore has two contributions: the power-law continuum (with a variable slope and normalisation) and the broad and narrow line-emitting components (each with its variable $T_{\rm e}$, $z$, $FWHM$, $E(B-V)$, and dust-corrected fluxes of all independent lines: \oiilowa, \neiiia, \hb, \oiiib, \heib). The best-fit continuum and line properties are presented in Table \ref{specfittable} while the resulting spectra are shown in Figure \ref{spec_1}.

\begin{table}
\caption{Best-fit continuum and observed (i.e., not dust corrected) line properties of each component, as derived through spectral fits. 
The best-fit continuum slope is defined as the power law slope of $F_{\lambda}$, and the flux density at $\lambda_{obs}=4\mu$m is given in units of $[\rm 10^{-20}erg\,s^{-1}\,cm^{-2}\,\angstrom^{-1}]$. For the narrow and broad component, we list the intrinsic (LSF-corrected) FWHM (in units of [km\,s$^{-1}$]) and redshift. The best-fit fluxes of each line are given in units of $[\rm 10^{-20}erg\,s^{-1}\,cm^{-2}]$. We only list the fluxes of the independent lines, and note that the dependent line fluxes (\oiilowa, \neiiia, \oiiia, \oiiic, and multiple Balmer lines) are determined using the listed $T_{\rm e}$, E(B-V), and the assumed $n_{\rm e}=200$\,cm$^{-3}$ (see Section \ref{moddes}).}
\label{specfittable}
\begin{tabular}{cccccc}
 &  & E & W\\\hline
Continuum Slope &  & $-1.01\pm0.24$ & $-1.45\pm0.17$\\\hline
F($\lambda_{obs}=4\mu$m) &  & $0.357\pm0.007$ & $0.480\pm0.007$\\\hline
$T_{\rm e}$ [$10^4$K] & N & $1.34\pm0.22$ & $1.16\pm0.10$\\
 & B & $1.00\pm0.32$ & (1.2)\\\hline
E(B-V) & N & (0.2) & $0.38\pm0.07$\\
 & B & (0.2) & (0.2)\\\hline
FWHM & N & $156\pm13$ & $231\pm8$\\
 & B & $653\pm11$ & $698\pm26$\\\hline
z & N & $7.1526\pm0.0001$ & $7.1487\pm0.0001$\\
 & B & $7.1533\pm0.0001$ & $7.1551\pm0.0007$\\\hline
F$_{\rm [OII]\lambda3727}$ & N & $86\pm10$ & $224\pm8$\\
 & B & $273\pm13$ & $62\pm9$\\
 & Total & $359\pm17$ & $286\pm12$ & \\\hline
F$_{\rm NeIII\lambda3968}$ & N & $210\pm18$ & $174\pm7$\\
 & B & $135\pm24$ & (0)\\
 & Total & $344\pm30$ & $174\pm7$\\\hline
F$_{\rm H\beta}$ & N & $273\pm13$ & $324\pm8$\\
 & B & $432\pm17$ & $104\pm9$\\
 & Total & $705\pm22$ & $428\pm12$\\\hline
F$_{\rm [OIII]\lambda5007}$ & N & $2472\pm98$ & $2318\pm57$\\
 & B & $3133\pm97$ & $599\pm59$\\
 & Total & $5605\pm138$ & $2917\pm82$\\\hline
F$_{\rm HeI\lambda5875}$ & N & $32\pm12$ & $46\pm5$\\
 & B & $75\pm20$ & (0)\\
 & Total & $107\pm23$ & $46\pm5$\\\hline

\end{tabular}
\end{table}

\subsubsection{Results and further measurables}\label{furmea}

Each spectrum features strong line emission (Figure \ref{spec_1}), with best-fit $T_{\rm e}$ values within $1\sigma$ of the fiducial value of $1.2\times10^4$\,K (Table \ref{specfittable}), comparable to conditions in other high-redshift galaxies (e.g., \citealt{hsai23,hu24,torr24}). From all of the regions and components, only the narrow component of component `W' features dust attenuation higher than the fiducial value.

The residuals around \oiiib (lower panels of Figure \ref{spec_1}) show sawtooth-like patterns. These residuals are only a few percent of the peak flux of \oiiib, and may imply a non-Gaussian LSF or emission profile. While these patterns are similar to those obtained when fitting a spectrum with a model that is lacking complexity (e.g., \citealt{gino20}), the addition of another free Gaussian component (representing either an outflow or an additional emitter) does not improve the fit. At first glance, they appear to be similar to the `wiggles' seen in other IFU studies (e.g., \citealt{pern23a}). However, these features are mostly seen from single-spaxel extractions that undersample the PSF, resulting in large-amplitude oscillations over a wide wavelength range. Since we extract spectra using larger apertures and observe artifacts that only appear within the FWHM of each line, they are likely not wiggles.

The use of a single continuum model (measured around \oiiib) for the full wavelength range results in low residuals for most of the lines explored here. One minor exception is that of \neiiib+\het in component `W', where the continuum appears to be overestimated. Because the continuum of neighbouring lines (\hz and \hd) are well-estimated, this may reflect a calibration issue in this wavelength range (e.g., unflagged cosmic ray hits or improper background subtraction). But because the line fluxes of \neiiib and \het are estimated using other line fluxes and ISM conditions, this does not affect any of our findings.

We use the best-fit line fluxes to calculate standard line ratios: 
\begin{itemize}
\item O32$\equiv$\oiiib/\oiilow
\item R3$\equiv$\oiiib/\hb
\item R2$\equiv$\oiilow/\hb
\item R23$\equiv$(\oiiiab+\oiilow)/\hb
\item Ne3O2$\equiv$\neiiia/\oiilow 
\end{itemize}
All of these (with the exception of Ne3O2) are then used to calculate the gas-phase metallicity using the strong-line diagnostics calibrated by \citet[][see their table 2 and equation 1]{curt20}. The resulting values are listed in Table \ref{specfittable2}. We find total gas-phase metallicities of $(Z_{\rm g}/Z_{\odot})\sim 0.2-0.3$ for each region, in general agreement with the values derived by \citet{suga24} using optical-FIR diagnostics (i.e., $Z_{\rm g}/Z_{\odot}\simeq0.2$).

Next, we infer the star formation rate (SFR) using the best-fit dust-corrected \hb flux. This is converted to a \ha flux using the extinction-free ratio of \ha/\hb (as derived with \pyneb assuming case B recombination). The conversion of \citet{redd18}, which is appropriate for the metallicity of this system ($Z_{\rm g}/Z_{\odot}\sim 0.28$), is then used to convert the estimated \ha flux to $SFR$. When calculating the uncertainty of this value, we include the errors on $T_{\rm e}$, \hb flux, and $E(B-V)$ (where a minimum uncertainty of 0.1 is assumed).

\begin{table}
\caption{Best-fit line ratios, gas-phase metallicities (\citealt{curt20}), \hb-based SFRs, and ionisation parameter $U$ (see Section \ref{furmea}) of each region. All values are dust corrected using the best-fit $E(B-V)$ values listed in Table \ref{specfittable}.}
\label{specfittable2}
\begin{tabular}{cccc}
 &  & E & W\\ \hline
O32 & N & $9.41\pm1.29$ & $2.68\pm0.13$\\
 & B & $3.77\pm0.23$ & $3.18\pm0.61$\\
 & Total & $5.12\pm0.29$ & $2.73\pm0.13$\\\hline
R3 & N & $8.84\pm0.56$ & $6.82\pm0.23$\\
 & B & $7.08\pm0.36$ & $5.64\pm0.73$\\
 & Total & $7.76\pm0.31$ & $6.67\pm0.22$\\\hline
R2 & N & $0.94\pm0.13$ & $2.54\pm0.12$\\
 & B & $1.88\pm0.12$ & $1.78\pm0.33$\\
 & Total & $1.52\pm0.09$ & $2.44\pm0.11$\\\hline
R23 & N & $12.74\pm0.74$ & $11.65\pm0.34$\\
 & B & $11.33\pm0.52$ & $9.31\pm1.02$\\
 & Total & $11.88\pm0.43$ & $11.35\pm0.33$\\\hline
Ne3O2 & N & $1.00\pm0.16$ & $0.31\pm0.02$\\
 & B & $0.20\pm0.04$ & $-$\\
 & Total & $0.39\pm0.04$ & $0.28\pm0.02$\\\hline
$Z_{\rm g}$ [$Z_{\odot}$] & N & $0.18\pm0.03$ & $0.32\pm0.04$\\
 & B & $0.27\pm0.04$ & $0.30\pm0.05$\\
 & Total & $0.23\pm0.03$ & $0.32\pm0.04$\\\hline
$SFR_{\rm H\beta}$ & N & $39\pm19$ & $99\pm38$\\
 & B & $61\pm27$ & $13\pm6$\\
 & Total & $100\pm33$ & $113\pm38$\\\hline
$\log_{10}(U)$ & N & $-2.24\pm0.05$ & $-2.68\pm0.02$\\
 & B & $-2.56\pm0.02$ & $-2.62\pm0.07$\\
 & Total & $-2.45\pm0.02$ & $-2.67\pm0.02$\\\hline
\end{tabular}
\end{table}

The combination of apertures E and W yield a total $\rm SFR=213\pm50\,M_{\odot}\,yr^{-1}$, which is in agreement with the SED-based estimate of \citet[][$\rm SFR=200^{+82}_{-38}\,M_{\odot}\,yr^{-1}$]{hash19} as well as the instantaneous ($\rm SFR_{0}=225^{+71}_{-56}\,M_{\odot}\,yr^{-1}$) and 10\,Myr-averaged ($\rm SFR_{10Myr}=207^{+65}_{-51}\,M_{\odot}\,yr^{-1}$) values from \citet{suga24}. Since $\rm SFR_{H\alpha}$ traces star formation over the past 5\,Myr (e.g., \citealt{flor21,tacc22}), this agreement is expected.  The fact that our SFR (from only the two core galaxies) is comparable to the other values (which were derived from the entire field) suggests that the majority of the star formation activity originates from the core galaxies rather than the diffuse emission.

We are also able to determine if this source lies on the star forming main sequence (SFMS). By combining the best-fit stellar mass of the system (\lmstar$ = 9.78^{+0.16}_{-0.18}$; \citealt{suga24}) with the framework of \citet{spea14}, the expected SFMS-based SFR of this source is $\rm 40^{+50}_{-22}\,M_{\odot}\,yr^{-1}$. The framework of \citet{meri25} yields a similar expected value of $\rm SFR=47^{+42}_{-22}\,M_{\odot}\,yr^{-1}$. The fact that we find a much higher SFR implies that \bfourteen is undergoing a starburst episode, in agreement with \citet{hash19}. This is also true for each core independently.

The O32 ratio has been shown to be an excellent tracer of the ionisation parameter \textit{U} (e.g., \citealt{naga06,papo22}). Following other recent works (e.g., \citealt{wits21,boye24,zamo24}), we convert the derived O32 values to \textit{U} using the relation of \citet{diaz00}:
\begin{equation}
    \log_{10}(U) = 0.80 \log_{10}\left({\rm O32}\right)-3.02
\end{equation}
which was derived using single star ionisation models. This yields values of $-2.7 \lesssim \log_{10}(U) \lesssim -2.2$ for the two core galaxies. Thus, the regions of \bfourteen show similar ionisation parameters as galaxies in the $2.7<z<6.3$ sample of \citet{redd23} and the lower redshift ($1.1<z<2.3$) sample of \citet{papo22}. 

\subsection{Spaxel-by-spaxel fits}\label{spaxelbyspaxel}
We can exploit the power of the NIRSpec IFU to extract physical properties on a spatially resolved basis by fitting combined line and continuum models to the spectra of each spaxel. This is enhanced through direct comparison to morpho-kinematic maps of ALMA data, which we create in an identical fashion.

\subsubsection{NIRSpec IFU map creation}\label{nmapc}
Due to the lower S/N of the spaxel-based spectra, we adopt a streamlined version of the model and fitting procedure of Section \ref{intspec} to investigate the spatial distribution of emission and ISM properties. First, we extract the spectrum from each spaxel and isolate the wavelength range spanning \oiilow to \oiiib\footnote{This wavelength range excludes \heib, which is too faint to spatially map.}. Instead of the multi-stage approach of Section \ref{intspec}, we fit the spectrum of each spaxel using a single model consisting of a power-law continuum and narrow and broad line contributions from \oiilow, \neiiiab, Balmer lines (\hz, \hep, \hd, \hz, \hg, \hb), \oiiic, and \oiiiab. The redshifts and intrinsic linewidths of the narrow and broad component of each line are set to be equal. 

While we explore a fit with a fiducial $n_{\rm e}$ and free $E(B-V)$ and $T_{\rm e}$ (as in the multi-stage model), we find that the S/N is too low to constrain these conditions. Instead, we fix $E(B-V)=0.2$ and $T_{\rm e}=1.2\times10^4$\,K throughout. By using the ISM conditions and \textlcsc{pyneb}, we lower the degrees of freedom in the model by predicting line ratios. Thus, the free variables for each spectrum are the continuum properties (slope and normalization) and independent fluxes ($F_{\rm [OII]\lambda3726}$, $F_{\rm [NeIII]\lambda3869}$, $F_{\rm H\beta}$, and $F_{\rm [OIII]\lambda5007}$) of the narrow and broad components of each line. As in Section \ref{moddes}, we reject components of lines with signal-to-noise levels of $<3\sigma$.

The resulting total intensity maps are included in Figure \ref{balmermaps}. 
Note that because of our model assumptions, the corresponding map of \neiiib is identical to that of \neiiia, but with a constant scaling factor (see Appendix \ref{pnapp}). Similarly, each of the other Balmer lines can be found from the \hb distribution, and \oiiia and \oiiic can be found from the \oiiib distribution.

Our fits also allow us to explore the kinematics of the field by extracting parameters from each spaxel. This is done by considering the best-fit \oiiib model, converting it to a cumulative distribution function (CDF), and calculating multiple non-parametric $v_{N}$ values (i.e., velocities at which the CDF reaches $N\%$). In each case we assume a fiducial zero-velocity redshift of $z=7.1520$. Figure \ref{velfig} shows $v_{50}$ (the mean velocity) and the asymmetry ($|v_{50}-v_{10}|-|v_{50}-v_{90}|$), while Figure \ref{widfig} shows $w_{80}\equiv v_{90}-v_{10}$ (a measure of linewidth). We note that in the case that the spectrum of a spaxel is best fit by a single component (either narrow or broad), the resulting asymmetry is zero.

\subsubsection{ALMA map creation}\label{almamapsec}

Previous analyses of the ALMA data for this source created moment maps (\citealt{hash19,hash23,suga21,suga24}), which are non-parametric measures of the total intensity (moment 0), intensity-weighted mean velocity (moment 1), and velocity dispersion (moment 2)\footnote{See CASA toolkit manual for more details: \url{https://www.aoc.nrao.edu/~kgolap/casa_trunk_docs/CasaRef/image.moments.html}}. Here, we instead use Gaussian models to fit the ALMA data on a spaxel-by-spaxel basis, resulting in morpho-kinematic maps that may be directly compared to our NIRSpec maps.

\begin{figure}
\centering
\includegraphics[width=0.5\textwidth]{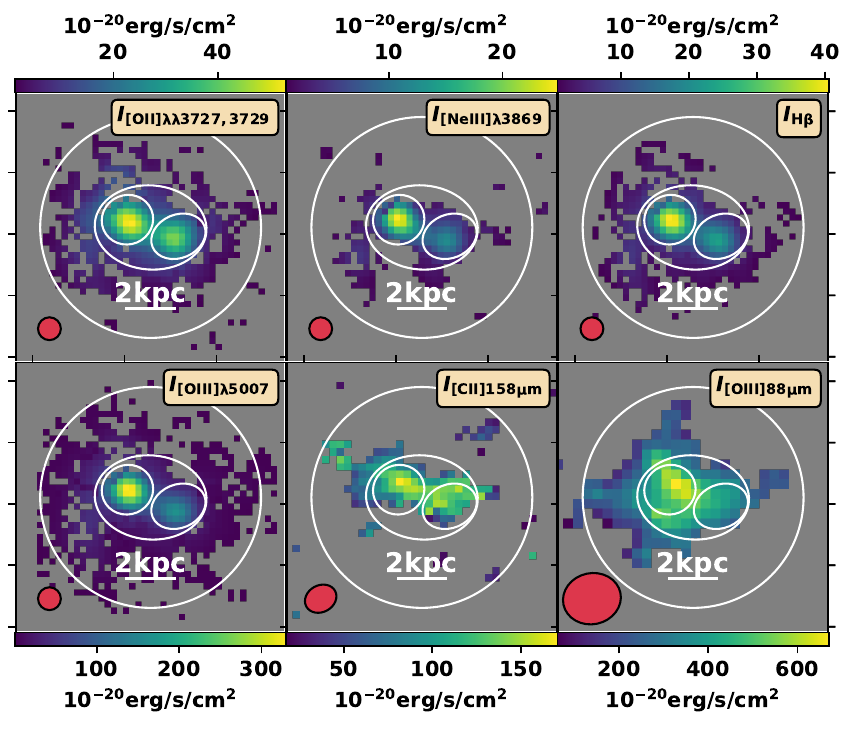}
\caption{Integrated total fluxes of emission lines, as derived through spaxel-by-spaxel fit. Only $>3\sigma$ fluxes are shown. Each panel displays a field of view of $2.2''\times2.2''$ centred on 10h01m40.69s +1$^{\circ}$54$'$52.55$''$. A physical scale bar of 2\,kpc scale bar is included in each panel.}
\label{balmermaps}
\end{figure}

Because each ALMA data cube only contains a single line (\cii or \oiiialma), the fitting process is much simpler than that of Section \ref{intspec} or \ref{spaxelbyspaxel}. First, we construct an error spectrum for each continuum-subtracted data cube by taking the RMS noise level of each spectral channel. Considering the possibility that each line may be fit by two Gaussian components (i.e., narrow and broad), the spectrum extracted from each spaxel is fit with these two Gaussians and an underlying first-order polynomial continuum using \textlcsc{lmfit}. However, this approach only reveals a few spaxels where a two-Gaussian model returns a better fit (i.e., a lower reduced $\chi^2$) in either cube. Because these potential areas are smaller than a synthesised beam and are not aligned with the broad emission found in \oiiib emission, they are likely artifacts (e.g., incomplete continuum subtraction, data combination, imaging). Thus, we adopt a single-Gaussian model. The resulting total intensity, $v_{\rm 50}$, and $w_{\rm 80}$ maps are included in Figures \ref{balmermaps}, \ref{velfig}, and \ref{widfig}, respectively.

\subsubsection{Distribution of emission}\label{doe}

All of the rest-optical lines feature similar morphologies, with emission focused in the cores (with the E nucleus being brighter), and weaker diffuse emission. The rest-FIR lines are qualitatively similar. The ALMA and JWST lines are further compared in Section \ref{almajwstsec}.

\begin{figure}
\centering
\includegraphics[width=0.5\textwidth]{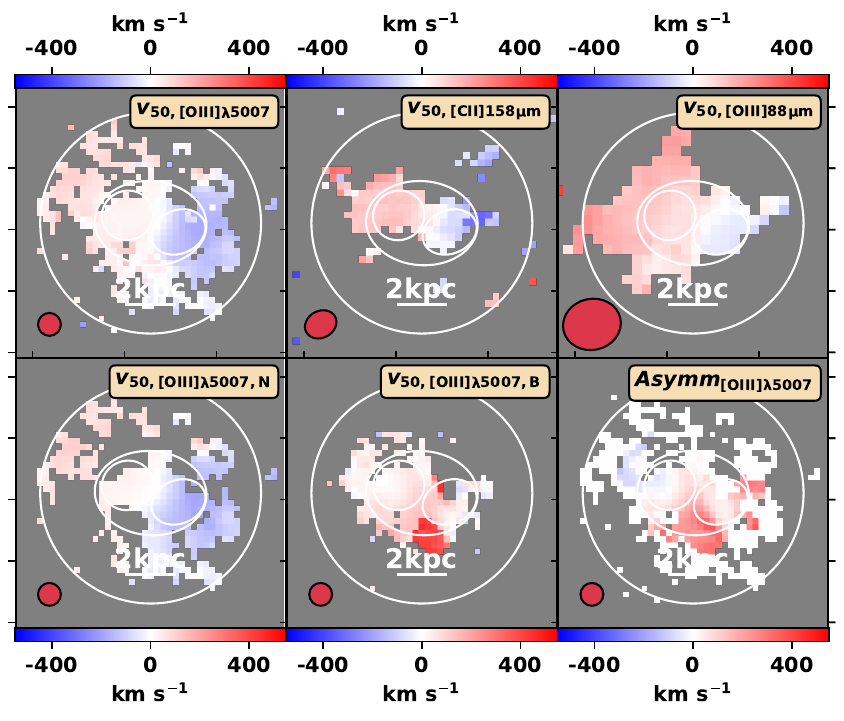}
\caption{Line of sight velocity maps, as derived through spaxel-by-spaxel fits. The top row includes maps of $v_{50}$ (i.e., the velocity at which each line reaches $50\%$ of its total flux) for \oiiib, \cii, and \oiiialma. Maps of $v_{50}$ for the narrow and broad component of \oiiib are shown in the first two panels of the lower row. The bottom right panel presents the best-fit asymmetry of \oiiib (see Section \ref{nmapc} for definition). Each panel displays a field of view of $2.2''\times2.2''$ centred on 10h01m40.69s +1$^{\circ}$54$'$52.55$''$. A physical scale bar of 2\,kpc scale bar is included in each panel. For each, we show the PSF as a red ellipse to the lower left. North is up and east is to the left.}
\label{velfig}
\end{figure}

We find that \oiiib, \cii, and \oiiialma feature similar distributions of $v_{\rm 50}$ (Figure \ref{velfig}), with a strong east-west velocity gradient. There are slight differences between the maps, which may be ascribed to differences in sensitivity and resolution (e.g., beam smearing, spatial binning). As seen in other works (e.g., \citealt{teli24}), \cii and \oiiialma show similar line-of-sight velocity maps. Our $v_{\rm 50}$ maps of the ALMA-detected lines are similar to the moment 1 maps created by \citet{hash19}, but differ slightly due to differences in imaging (e.g., spaxel size) and velocity measurement method (i.e., Gaussian fit vs. moment 1). While such a symmetric velocity field could be interpreted as a signature of rotation, the disparate properties of each core (see Table \ref{specfittable2}, Figure \ref{diagmaps}) suggest a major merger instead (see Section \ref{naturecomp} for further discussion).

The narrow and broad components of \oiiib show different spatial distributions (see Appendix \ref{NBAPP} for more details) and velocity offsets (Figure \ref{velfig}). Due to this, there is a region of positive asymmetry to the southeast of core W that indicates a more powerful red broad component. Indeed, the channel maps of both \oiiia and \oiiib contain low-level redshifted emission in this area, which may indicate a tidal feature or outflow.

\begin{figure}
\centering
\includegraphics[width=0.5\textwidth]{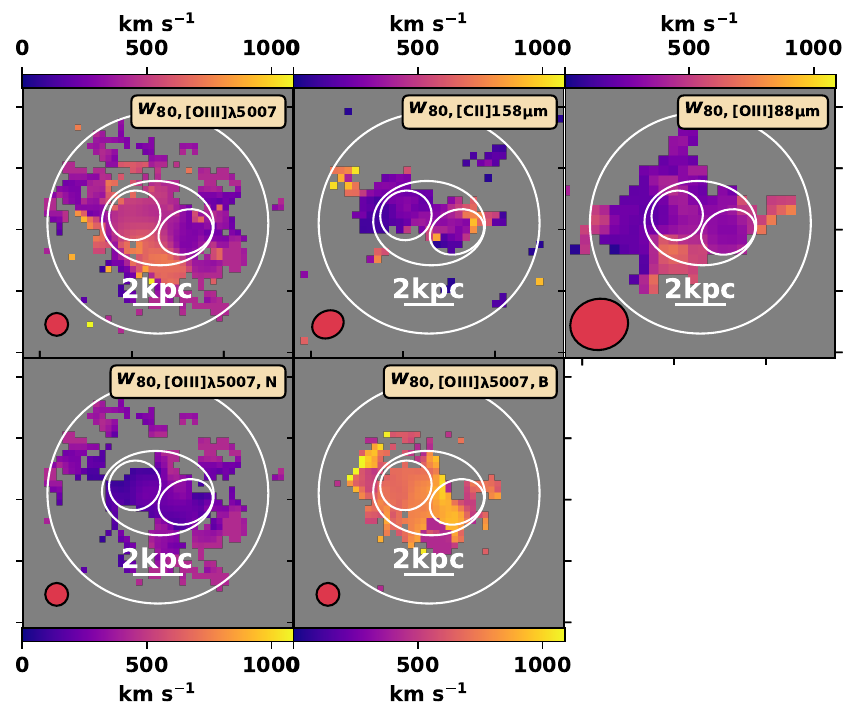}
\caption{Maps of $w_{80}$ (i.e., the difference in velocity between the points at which each line reaches $10\%$ and $90\%$ of its total flux), as derived through spaxel-by-spaxel fits for \oiiib, \cii, and \oiiialma. Maps of $w_{80}$ for the narrow and broad component of \oiiib are shown in the lower row. Each panel displays a field of view of $2.2''\times2.2''$ centred on 10h01m40.69s +1$^{\circ}$54$'$52.55$''$. A physical scale bar of 2\,kpc scale bar is included in each panel. For each, we show the PSF as a red ellipse to the lower left. North is up and east is to the left.}
\label{widfig}
\end{figure}

We are also able to calculate $w_{80}$ for each line (Figure \ref{widfig}). Core E has a higher mean $w_{80}$ ($\sim470$\,km\,s$^{-1}$) than core W ($\sim370$\,km\,s$^{-1}$), which is primarily caused by the stronger broad component in core E. The $w_{80}$ map of \oiiialma shows a region with elevated velocity dispersion south of core E which slightly overlaps with the broad \oiiib emission, but the two cores show similar $w_{\rm 80,[OIII]88\mu m}$. 

The two ALMA $v_{50}$ maps are more similar to that of the narrow component of \oiiib, suggesting that the gas emitting broad \oiiib is not detected in \cii or \oiiialma. We note that the JWST/MIRI observations did not detect broad \ha emission, which appears to be at odds with our detection of narrow and broad Balmer emission. However, this is likely simply due to lower sensitivity (see discussion in \citealt{prie25}).

\subsubsection{Further measurables}
\begin{figure}
\centering
\includegraphics[width=0.5\textwidth]{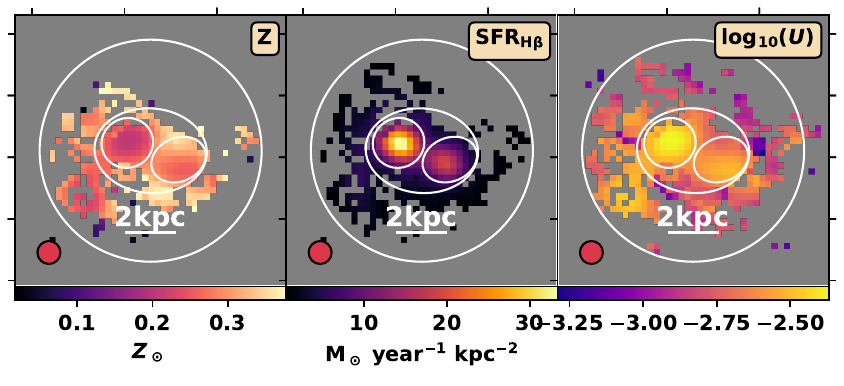}
\caption{Line ratio diagnostics, as derived through spaxel-by-spaxel fit. Each panel displays a field of view of $2.2''\times2.2''$ centred on 10h01m40.69s +1$^{\circ}$54$'$52.55$''$. A physical scale bar of 2\,kpc scale bar is included in each panel. For each, we show the PSF as a red ellipse to the lower left. North is up and east is to the left.}
\label{diagmaps}
\end{figure}

The best-fit total line maps (i.e., narrow + broad flux; Figure \ref{diagmaps}) are used to create maps of gas-phase metallicity, SFR, and $U$ (see Section \ref{intspec} for ratio definitions and diagnostic information). The resulting maps for the combined best-fit models are shown in Figure \ref{diagmaps}.

Each map yields results that broadly agree with the integrated spectral analysis of Section \ref{furmea}. That is, the W core features higher metallicity and lower $\log_{10}(U)$. In addition, a high-ionization and low-Z arc (which is detected in \oiilow, \hb, and \oiiib; Figure \ref{balmermaps}) is visible to the southeast. The $SFR_{\rm H\beta}$ map (which is a scaled version of the \hb map of Figure \ref{balmermaps}) is less illustrative, as a uniform dust correction has been applied. Because of this, the SFR density in this map may be treated as a minimum value.

\section{Discussion}\label{discsec}

\subsection{Line ratio diagrams}\label{LRD}
Due to the wavelength coverage of our data, we do not have access to \niiab or \siiab, and cannot utilise the standard [NII]-BPT or [SII]-VO87 diagrams (\citealt{bald81,veil87}) to search for evidence of AGN excitation. But the wealth of detected emission lines (see Table \ref{specfittable}) allows us to explore the nature of this field using other diagnostics: O32-R23, R3-R2, and Ne3O2 (see definitions in Section \ref{intspec}).

\begin{figure*}
\centering
\includegraphics[width=\textwidth]{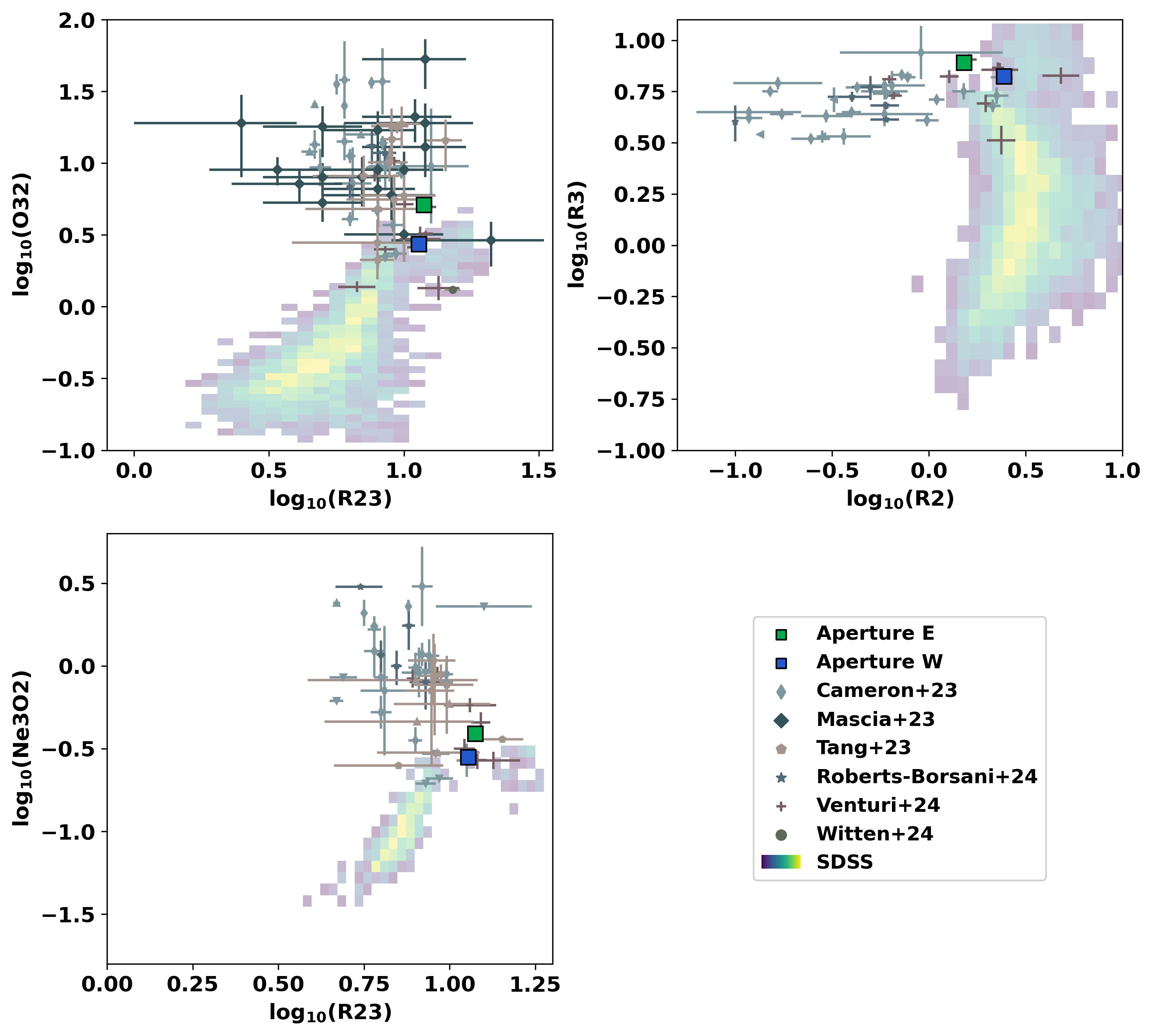}
\caption{Distribution of line ratios for the apertures E and W: O32-R23 (top left), R3-R2 (top right), and Ne3O2-R23 (bottom left). For comparison we also show the dust-corrected line ratios of low-redshift galaxies from BOSS (\citealt{thom13}) and higher redshift galaxies previously observed with JWST/NIRSpec ($5.5\lesssim z\lesssim9.5$, \citealt{came23}; $4.6\lesssim z\lesssim7.9$, \citealt{masc23}; $6.9\lesssim z\lesssim 9.0$, \citealt{tang23}; $5 \lesssim z\lesssim9$, \citealt{robe24}; $6.4\lesssim z\lesssim 7.9$, \citealt{vent24}; and $z=7.88$, \citealt{witt24}).}
\label{ratplot}
\end{figure*}

The ratios of the E and W cores of \bfourteen are shown in Figure \ref{ratplot}. For comparison we also show the dust-corrected line ratios of low-redshift galaxies from Sloan Digital Sky Survey-III/Baryonic Oscillation Spectroscopic Survey
(BOSS; \citealt{daws13,thom13}), $5.5\lesssim z\lesssim9.5$ galaxies from the JADES survey \citep{came23}, $4.6\lesssim z\lesssim7.9$ galaxies from \citet{masc23}, $6.9\lesssim z\lesssim 9.0$ galaxies from the CEERS survey \citep{tang23}, A2744-YD4 ($z=7.88$; \citealt{witt24}), and the $5 \lesssim z\lesssim9$ stacks created by \citet{robe24}. Each of the line ratios from high-redshift galaxies has either been corrected for dust attenuation or has been measured for galaxies with no significant attenuation. Generally, we find that the different components of \bfourteen lie within the scatter of the JWST-observed galaxies, at the upper edge of the scatter of the low-redshift galaxies. Below we discuss the physical interpretation of this line ratio distribution.

All of these ratios are dependent on the ionisation parameter \textit{U} and metallicity $Z_{\rm g}$ (e.g., \citealt{kewl02,shir14,came23}). In general, R23, R2, and R3 are stronger tracers of $Z_{\rm g}$ (e.g., \citealt{stei16,curt20}), while Ne3O2 and O32 are more dependent on $U$ (e.g., \citealt{pens90,leve14}). The low-redshift galaxies show a positive correlation in each diagnostic plot, while the high-redshift galaxies lie at higher O32 and Ne3O2 (suggesting higher $U$, \citealt{came23}). Both regions of \bfourteen lie near the intersection of low-redshift and previous JWST-observed galaxies.

The metallicity of each region of \bfourteen was found to be $\sim0.2-0.3\,Z_{\odot}$ (Section \ref{intspec}), allowing us to use these diagrams to compare ionisation states. Since the \bfourteen regions have low O32 and Ne3O2 compared to the other high-redshift sources, this suggests that our sources have relatively low ionisation parameters. The high $Z_{\rm g}$ and low $U$ (compared to other high-redshift galaxies) suggests that \bfourteen could represent a system of evolved galaxies in the early Universe where significant star formation has occurred (resulting in a higher $M_*$ and $Z_{\rm g}$). Indeed, they lie at the high-$Z_{\rm g}$ and high-$U$ edge of low-redshift galaxies, despite having much less time to form stars and enrich their environments.

\subsection{ALMA-JWST comparison}\label{almajwstsec}
While JWST is a powerful tool to explore the characteristics of high-redshift galaxies, it also features a strong synergy with ALMA that has only recently begun to be explored. This includes the possibility of detecting Lyman continuum leaking galaxies by comparing \cii and \oiiib (e.g., \citealt{katz20,ura23}), using JWST-based gas-phase metallicities and ALMA-based CO observations to derive molecular gas masses (e.g., \citealt{nara12}), and comparing \oiiib/\oiiialma to place tight constraints on electron temperature (e.g., \citealt{stia23}). 

\subsubsection{Lyman continuum leakage}
The escape of Lyman continuum photons is a key process in the study of how the Universe was reionised, but due to the opacity of the IGM, this quantity is usually explored using indirect tracers (e.g., \citealt{izot18,chis22,masc23}). A novel approach was proposed by \citet{katz20}, who used high-resolution cosmological simulations to train a logistic regression model to predict whether a galaxy has a high Lyman continuum escape fraction ($f_{esc}^{LyC}>10\%$; `Lyman leakers') using multiple strong rest-UV and FIR lines. 

Before the advent of JWST, the most easily applied diagnostic was between two FIR lines that were observable with ALMA ($\rm [CII]158\mu m$ and $\rm [OIII]88\mu m$). From this diagnostic, \citet{katz20} predict that \bfourteen is likely a non-leaker. However, the strongest predictor of high $f_{esc}^{LyC}$ was found to be between \oiiib and \cii. Since \cii can trace neutral gas (e.g., \citealt{vizg22}, but see also e.g., \citealt{zane18,madd20}) and \oiiib traces ionised gas, a high value of this ratio could indicate a higher level of ionised gas and thus more avenues of escape for Lyman continuum photons. Here, we examine this ratio in \bfourteen.

Because the astrometric uncertainty of the ALMA images are comparable to the spaxel size of our JWST/NIRSpec data (see Appendix \ref{astro3}), we do not consider the spatial distribution of line ratios from ALMA and JWST. Instead, we consider the total line flux ratios of each using the `Total' aperture (Table \ref{aper_table}). 

We find $\log_{10}(L_{\rm [CII]158\mu m}\rm\,[erg\,s^{-1}])=42.64\pm0.06$, and $\log_{10}(L_{\rm [OIII]\lambda5007}\rm\,[erg\,s^{-1}])= 43.65\pm0.02$ (with no dust extinction correction). Despite this relatively high value of $\log_{10}(L_{\rm [OIII]\lambda5007})$, these integrated values place \bfourteen in the `non-leakers' region of the \citet{katz20} diagram, in agreement with the classification based on the \cii-\oiiialma diagnostic. While the application of a dust correction ($E(B-V)>0$) will result in a higher \oiiib luminosity and push \bfourteen towards the leaker regime, a large correction ($E(B-V)>0.68$) is required for a leaker probability of $>50\%$.

\subsubsection{Electron density}
A previous exploration of \oiiiab/\oiiialma was performed by \citet{suga24}, who combined ALMA and JWST/NIRCam observations of \bfourteen to explore how the electron density varies across this source. While these previous observations featured higher spatial resolution, our JWST/NIRSpec data allow us to directly characterise the rest-optical line and continuum emission. We repeat this analysis here using our NIRSpec IFU data. Similarly to above, we do not investigate the spatial distribution of this ratio due to astrometric alignment issues, but use the fluxes found using the `Total' aperture and \pyneb to constrain the density.

The ratio of the best-fit \oiiib and \oiiialma fluxes ($2.6\pm0.3$) falls outside the ratios explored by \pyneb ($\gtrsim3.8$). While this could be interpreted as very low density ($\log_{10}(n_{\rm e})<10^1$\,cm$^{-3}$), we note that we may recover our fiducial density ($\log_{10}(n_{\rm e})=10^2$\,cm$^{-3}$) by adopting a reasonable $E(B-V)\gtrsim0.09$. Since the field contains regions of both high (e.g., `W') and low dust extinction (`E'), this extinction is reasonable. While it is not straightforward to derive a global dust correction, this \oiiib/\oiiialma ratio suggests that \bfourteen does not feature areas of extremely high density (which would require a very high ratio).

To investigate this further, we require higher spectral resolution observations of \oiilow (i.e., JWST/NIRSpec G395H/F290LP) and/or higher spatial resolution and sensitivity observations of \oiiialma (i.e., ALMA).

\subsection{Presence of dust}\label{pod}

The fact that this source is detected in FIR continuum emission with ALMA (e.g., \citealt{hash19}) implies that there is a significant dust reservoir in the field. Through a detailed analysis of the dust SED (which has three detections and one non-detection), \citet{suga21} find a total dust mass of $\sim10^{6-8}\,\rm M_{\odot}$ (in agreement with previous results; \citealt{hash19}). The recent SED modelling of \citet{suga24} found a stellar continuum dust attenuation of $A_{\rm V}^*=0.78^{+0.07}_{-0.06}$ (corresponding to ionised gas $E(B-V)\sim0.4$; \citealt{calz00}) for the full source. Our JWST/NIRSpec spectra are well fit by models with low $E(B-V)=0.2-0.4$, but we lack the sensitivity and wavelength coverage necessary to estimate dust attenuation further (See Appendix \ref{lri} for further discussion of this).

\begin{figure*}
\centering
\includegraphics[width=\textwidth]{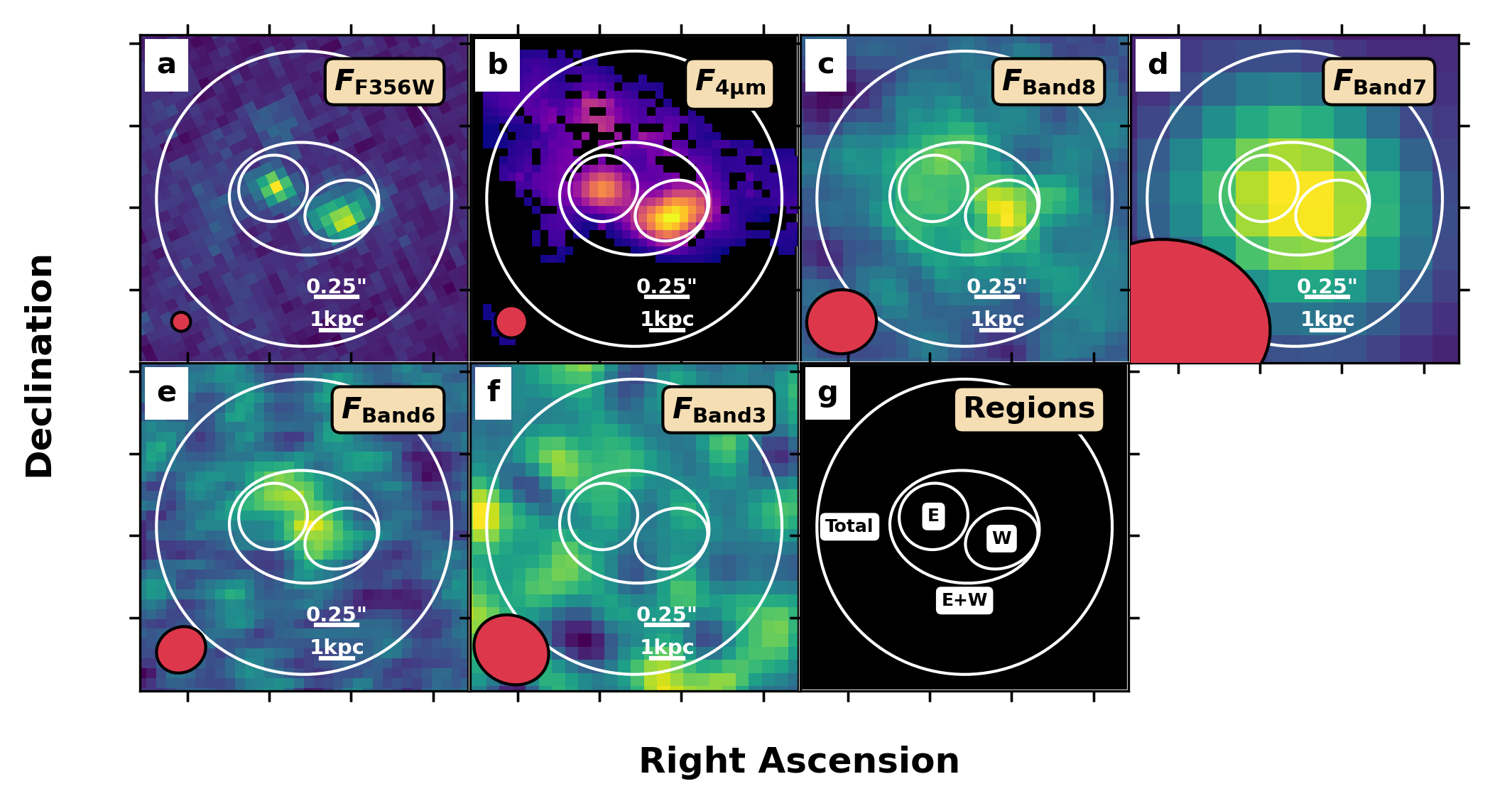}
\caption{Gallery of multi-wavelength continuum data (see Table \ref{MWtable} for details) in order of increasing $\lambda_{obs}$. Each panel displays a field of view of $1.8''\times1.8''$ ($\sim9.3\times9.3$\,kpc$^2$), centred on 10h01m40.6858s +1$^{\circ}$54$'$52.553$''$. An angular and physical scale is shown in each panel. {The regions of study (which are adopted from \citealt{prie25}) are shown in panel h}. For each, we show the PSF as a red ellipse to the lower left. North is up and east is to the left. The JWST/NIRSpec continuum map (panel b; $\lambda_{obs}=4\,\mu$m), which was derived through spaxel-by-spaxel fit (see Section \ref{spaxelbyspaxel} for details), is displayed in a different colourmap for emphasis.}
\label{contmap}
\end{figure*}

Most of the previous studies of the FIR emission used the spatially resolved band 6 and 8 data to construct SEDs for the E and W cores, which were then analysed with different models (i.e., a blackbody model, \citealt{hash19}; Bagpipes, \citealt{carn18bagpipes,suga24}; or CIGALE, \citealt{noll09,boqu19,prie25}). But since each used slightly different spatial apertures (which were centred on line emission rather than continuum emission), we consider the dust morphology of this field (see Figure \ref{contmap}). 

Our analysis of the rest-UV and rest-optical line emission shows that these originate primarily from two bright cores, with weaker diffuse emission. Emission from the rest-FIR lines \cii and \oiiialma also is focused in the two core regions. Similarly, the Band 8 continuum emission (underlying \oiiialma and tracing the peak of the dust SED) originates from core W and features a significant diffuse component. But the Band 6 emission (underlying \cii) is focused in an arc-like feature between the two core regions ($\sim0.2''$ from the centre of either core). This morphology was also reported by previous works (\citealt{hash19,suga24}), and thus is likely not an artefact introduced by our calibration or imaging processes. This offset is also larger than the positional uncertainty of the related \cii map ($\sim0.03''$; Appendix \ref{astro3}). Neither the band 7 nor band 3 maps contain morphological information due to lack of spatial resolution and lack of detection, respectively. Because the emission in each ALMA band is not coincident, it possibly originates from different dust reservoirs: one dust reservoir located in core W (see Band 8 image in panel d of Figure \ref{contmap}), as well as a separate dust reservoir between the cores (Band 6). Alternatively, dust temperature gradients have been observed at high redshift (e.g., \citealt{cali18,dong19,tsuk23}), which would allow for this change in morphology. Indeed, the previous SED fits found different dust temperatures in each core (e.g., \citealt{hash19,suga24}).

While a spatially resolved FIR SED analysis would unlock a deeper understanding of the dust properties and distribution (e.g., \citealt{tsuk23}), it requires time-intensive observations and is currently only feasible for bright sources (e.g., starburst galaxies, QSO host galaxies). The lack of detection in band 3 and the low spatial resolution of band 7 means that such a spatially resolved analysis of the FIR emission would only feature two SED points, and would thus be biased by assumptions of flux distribution, dust properties, and/or the use of a FIR template. With the data available, we may only state that the \bfourteen field contains dust that is spatially offset from the bright core regions. This dust may have been expelled by stellar winds (e.g., \citealt{veil05}), or may be affected by the ongoing galaxy merger (e.g., \citealt{tamu23}).

\begin{table}
    \centering
    \begin{tabular}{c|ccc}
Instrument	&	Name	&	$\lambda_{obs}$	&	$\lambda_{rest}$	\\	
	&		&	[$\mu$m]	&	[$\mu$m]	\\	\hline
JWST/NIRCam Imaging	&	F356W	&	3.563	&	0.437	\\	
JWST/NIRSpec IFU	&	$4\,\mu$m	&	4.000	&	0.491	\\	
ALMA	&	Band8	&	740.228	&	90.803	\\	
ALMA	&	Band7	&	999.308	&	122.584	\\	
ALMA	&	Band6	&	1332.411	&	163.446	\\	
ALMA	&	Band3	&	3331.027	&	408.615	\\	
\end{tabular}
\caption{List of continuum data. For each dataset, we list a representative wavelength in the observed and rest frame (assuming $z=7.1520$): pivot wavelengths for each JWST imaging filter and average wavelengths for each ALMA band.}
\label{MWtable}
\end{table}

\subsection{Mass-metallicity relation}

It has been well established that the properties of most galaxies at a set redshift tend to scale with each other. Examples of this are seen in the observed relations between stellar mass and rotational velocity (i.e., the Tully-Fisher relation; \citealt{tull77}), SFR and $M_*$ (star-forming main sequence; \citealt{noes07}), and the surface densities of molecular and stellar mass (molecular gas main sequence; \citealt{lin19}). Here, we investigate the placement of \bfourteen on the relation between stellar mass and gas-phase metallicity, or the mass-metallicity relation (MZR; \citealt{lequ79}). 

\begin{figure*}
    \centering
    \includegraphics[width=\textwidth]{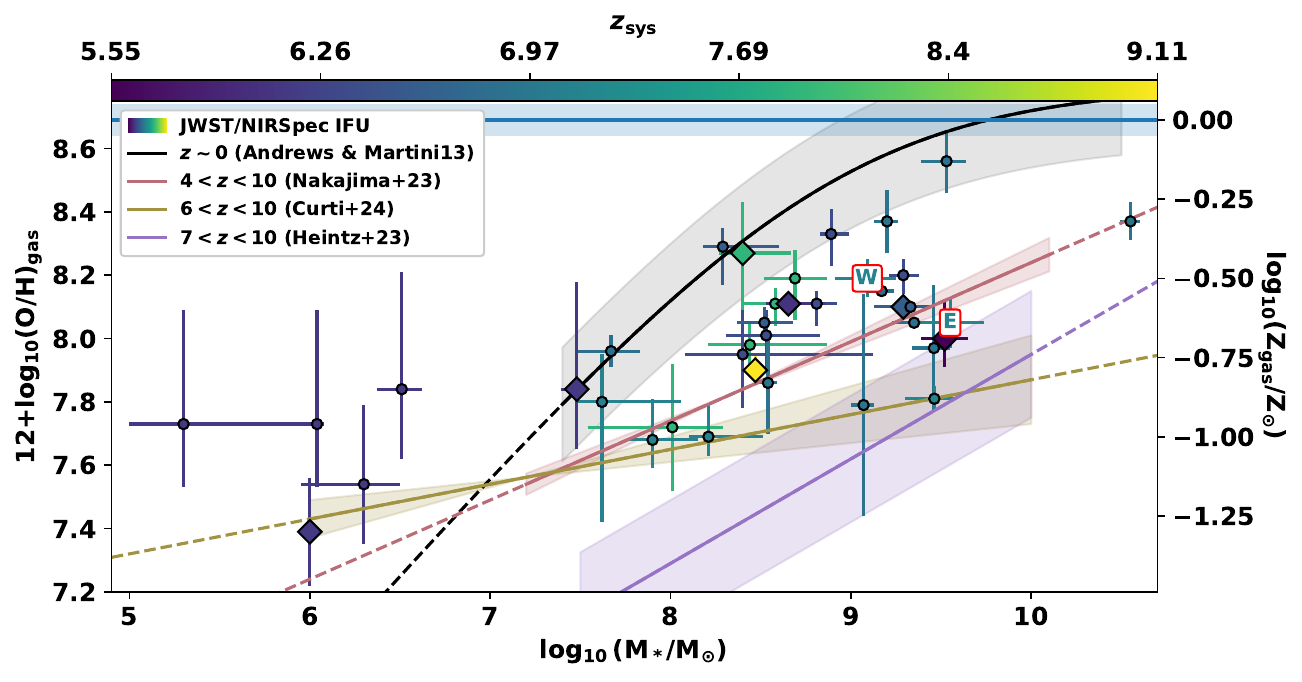}
    \caption{Gas-phase metallicities of $z>5$ galaxies observed with the JWST/NIRSpec IFU as a function of stellar mass (\citealt{uble23,uble24zs7,arri24,fuji24,ji24,marc24b,marc24a,mess24,scho24,vent24}). Each point is coloured by redshift. Values that represent entire fields rather than regions are shown as diamonds. We also show previously determined MZR fits from local studies (\citealt{andr13}) and high-redshift studies (\citealt{hein23,naka23,curt24}). Our values of core E and core W are shown by `'E' and `W', respectively.}
    \label{MZRfig}
\end{figure*}

To examine the location of \bfourteen with respect to the high-redshift MZR, we display $Z_{\rm g}$ (derived using the line diagnostics of \citealt{curt20}; Section \ref{furmea}) and $M_*$ (derived by a BAGPIPES fit to the observed SED; \citealt{suga24}) for core E and core W in Figure \ref{MZRfig}. These values are compared to existing best-fit MZR relations (\citealt{hein23,naka23,curt24}) as well as $z>5$ galaxies with $Z_{\rm g}$ and $M_*$ from JWST/NIRSpec IFU observations (\citealt{arri24,marc24b,marc24a,mess24,scho24,vent24}). We find that the observed galaxies lie above the MZR as derived by \citet{hein23} and \citet{curt24}, and are closer to the relation of \citet{naka23}. 

However, there are important caveats to this interpretation that must be discussed. Primarily, the IFU targets (which mainly originate from the GA-NIFS survey) are not a representative sample of galaxies, but are instead biased towards bright, well-studied objects (e.g., SPT0311-58-E with $\log_{10}(M_*/M_{\odot})=10.55^{+0.05}_{-0.06}$, \citealt{arri24}). While these observations result in the detection of smaller satellite galaxies (e.g., SPT0311-58-L1 with $\log_{10}(M_*/M_{\odot})=8.54\pm0.05$, \citealt{arri24}), we lack a more complete, unbiased dataset of IFU-detected galaxies in this redshift range ($6\lesssim z\lesssim9$). The galaxies at $\log_{10}(M_*/M_{\odot})<7$ originate from the work of \citet{mess24}, where gravitational magnification ($\mu\sim20$) allows for the characterisation of dwarf galaxies. The high-metallicity source is SPT0311-58-C3, which lies near the edge of the IFU field of view and thus may suffer from low sensitivity. As noted above, these works also use different methods to derive $Z_{\rm g}$ and $M_{*}$. A uniform analysis of NIRSpec IFU data will be performed in a future work.

We find that the cores of B14-65666 are among the most massive galaxies in our collected sample, but they still lie on an extrapolation of the \citet{naka23} MZR trend. This suggests that the masses of each core are not dominated by a large reservoir of pristine gas (which would result in a lower metallicity), but they are also unlikely to have lost large amounts of gas via outflows or tidal interactions (which would result in a higher metallicity). Instead, it appears that the cores of B14-65666 have formed stars and enriched their gas in a similar fashion to other high-redshift galaxies.

\subsection{Nature of core components}\label{naturecomp}
The two core galaxies of \bfourteen are closely associated, with a projected distance of $\sim0.43''$ (corresponding to $\sim2.2$\,kpc) and a velocity offset of $\sim175$\,km\,s$^{-1}$. Both cores appear to be undergoing a starburst episode (Section \ref{furmea}). Despite this, some of their properties reveal distinct natures. On one hand, core E appears to have a significant molecular reservoir (as suggested by its strong \cii emission; e.g. \citealt{zane18,madd20}), lower $Z_{\rm g}$, and higher ionisation parameter $U$. With a compact JWST/NIRCam morphology and high $M_*$, this core likely represents a galaxy undergoing a burst of star formation activity.

Core W is the opposite in multiple aspects: it has a lower $M_*$ and amount of molecular gas (i.e., weaker \cii), higher $Z_{\rm g}$, and lower $U$. As seen in Figure \ref{MZRfig}, this core lies above the high-$z$ MZR, putting it closer to more evolved, lower-$z$ populations. Thus, this core appears to be a galaxy that has used much of its molecular gas in star formation.

Figure \ref{velfig} shows that there is a clear velocity gradient between these two galaxies, with a broad, redshifted component between them. Such a velocity gradient could be interpreted as a sign of rotation, but this is unlikely here due to the presence of two clearly separated galaxies (i.e., core E and core W) with different properties. Since the line and continuum emission peak in one (or both) of the cores rather than between them (except for the ALMA band 6 continuum, see Section \ref{pod}), it is unlikely that this represents a single rotating galaxy, even accounting for the effects of dust obscuration.

Instead, it is likely that these cores represent merging galaxies that have undergone multiple close interactions in the past, as suggested by previous works (\citealt{hash19,suga24,prie25}). Such interactions would strip gas off of each galaxy (explaining the presence of the surrounding diffuse emission), and could boost or lower star formation in each source (e.g., \citealt{dima08,hors21,elli22}). The redshifted, broad line emission between the clumps (which is found here for the first time) could then represent stripped gas from a previous interaction. 

This behaviour has been observed in other systems, such as CR7 \citep{marc24b}: a $z=6.6$ field composed of three primary components, an extended \lya halo, and multiple minor components. A SED fit yielded SF histories (SFHs) for each component, revealing evidence for a past merger-induced starburst. SED fitting of \bfourteen by \citet{suga24} suggest that the SFH of this field is dominated by a starburst $\sim10$\,Myr prior to $z=7.1520$. Additional observations with JWST/NIRSpec (e.g., to observe \lya and enable further SED modelling) could be useful for characterising this field.

\section{Conclusions}\label{concs}
In this work, we have presented new JWST/NIRSpec IFU observations of the $z=7.1520$ galaxy \bfourteen as part of the GA-NIFS survey. Through combination with archival data from JWST/NIRCam and ALMA, we are able to explore the morpho-kinematics and ISM conditions on a spectral and spatial basis.

We confirm that the emission from this object primarily originates from two bright cores (`core E' and `core W'), surrounded by low-level diffuse emission. Spectra extracted from each core reveal significant emission of \oiilow, \neiiiab, \oiiiab, several Balmer lines (\hb, \hg, \hd, \hep, \hz, \het), \heib, \heic, and weak \oiiic. By using \pyneb to calculate line ratios for given electron conditions ($T_{\rm e}$ and $n_{\rm e}$), we are able to account for blended lines and the relatively low spectral resolution of the data ($R\sim1000$). By combining these data with previous ALMA \cii and \oiiialma data, we are able to characterise the field with unprecedented detail.

Using strong line diagnostics \citep{curt20}, we constrain the gas-phase metallicity of each core to be $\sim0.2-0.3\,Z_{\odot}$. The line ratios of each core lie within the scatter of other $z\sim 5.5-9.5$ galaxies observed with JWST/NIRSpec (\citealt{came23,tang23}), but with low O32 and Ne3O2 (i.e., low $U$) and high R23 and R2 (i.e., higher $Z_{\rm g}$). Since these properties are more similar to the high-ionisation tail of the distribution of low-redshift galaxies observed in BOSS (\citealt{thom13}), we propose that \bfourteen represents a system of currently starbursting galaxies in the early Universe, where significant star formation has already occurred.

Next, we maximise the potential of our three-dimensional dataset by performing spaxel-by-spaxel fits. This reveals that the emission of each JWST/NIRSpec- and ALMA-detected lines are concentrated in the two cores. By fitting the rest-UV and rest-optical lines with a broad and narrow component, we find a region of redshifted, broad emission between the two cores. This likely represents a tidal interaction between the two cores (or possibly an outflow). On the other hand, the narrow emission shows a clear east-west red-blue velocity gradient, which is also seen in \cii and \oiiialma.

The ALMA and JWST data are combined to test optical-FIR line diagnostics of the full field. These suggest that B14-65666 may not be a significant Lyman continuum leaker and is unlikely to feature a high $n_{\rm e}$ (but this depends on the adopted dust extinction). A comparison of the continuum and line maps suggests that \bfourteen may have a complex dust distribution.

The $M_*$ of \bfourteen (as derived by \citealt{suga24}) is combined with our precise estimate of $Z_{\rm g}$ to examine if this source lies on the high-redshift MZR. By comparing these values with other JWST/NIRSpec IFU-derived values and best-fit trends of representative galaxies, we find that cores of \bfourteen lie on the high-redshift MZR. This suggests the lack of a large molecular reservoir or gas expulsion via feedback, which would result in significant offset from the relation.

By combining multiple tracers (e.g., $Z_{\rm g}$, line morpho-kinematics), we determine that the two cores feature drastically different properties. Core W is a less massive, lower-metallicity galaxy that appears to have already used up much of its molecular gas reservoir in star formation (based on the less significant \cii peak). Core E features a higher $M_*$ and a higher \cii luminosity, suggesting higher potential for future star formation. When combined with the kinematics we have observed, the two cores of \bfourteen likely represent a massive merger of two disparate galaxies in the early Universe, which have created tidal features through their past interactions. 

\section*{Acknowledgements}
We thank the anonymous referee for their constructive feedback that has resulted in a stronger work.
GCJ and JS acknowledge support by the Science and Technology Facilities Council (STFC), by the ERC through Advanced Grant 695671 ``QUENCH'', and by the UKRI Frontier Research grant ``RISEandFALL.''
GCJ, AJB, and JC acknowledge funding from the ``FirstGalaxies'' Advanced Grant from the European Research Council (ERC) under the European Union’s Horizon 2020 research and innovation programme (Grant agreement No. 789056).
RB acknowledges support from an STFC Ernest Rutherford Fellowship (ST/T003596/1).
SA, MP, and BRP acknowledge support from grants PID2021-127718NB-I00, RYC2023-044853-I, and PID2024-159902NA-I00, funded by the Spanish Ministry of Science and Innovation/State Agency of Research (MCIN/AEI/10.13039/501100011033) and El Fondo Social Europeo Plus FSE+. 
SCa and GV acknowledge support from the European Union (ERC, WINGS, 101040227).
H\"{U} gratefully acknowledges support by the Isaac Newton Trust, the Kavli Foundation through a Newton-Kavli Junior Fellowship, and funding by the European Union (ERC APEX, 101164796). Views and opinions expressed are however those of the authors only and do not necessarily reflect those of the European Union or the European Research Council Executive Agency. Neither the European Union nor the granting authority can be held responsible for them.
GC acknowledges the support of the INAF Large Grant 2022 ``The metal circle: a new sharp view of the baryon cycle up to Cosmic Dawn with the latest generation IFU facilities''.
This paper makes use of the following ALMA data: ADS/JAO.ALMA\#2015.1.00540.S, ADS/JAO.ALMA\#2016.1.00954.S, ADS/JAO.ALMA\#2017.1.00190.S, ADS/JAO.ALMA\#2018.1.01673.S, ADS/JAO.ALMA\#2019.1.01491.S. ALMA is a partnership of ESO (representing its member states), NSF (USA) and NINS (Japan), together with NRC (Canada), MOST and ASIAA (Taiwan), and KASI (Republic of Korea), in cooperation with the Republic of Chile. The Joint ALMA Observatory is operated by ESO, AUI/NRAO and NAOJ.

\section*{Data Availability}
The JWST/NIRSpec data used in this research has been obtained within the NIRSpec-IFU GTO programme GA-NIFS (PID 1217) and are publicly available on the MAST archive, along with all analysed JWST/NIRCam data. Data presented in this work will be shared upon reasonable request to the corresponding author.

\bibliographystyle{mnras}
\bibliography{example}

\appendix

\section{PSF treatment}\label{psfapp}
The JWST/NIRSpec IFU PSF is known to vary strongly with wavelength (e.g, \citealt{deug23a}). Recent works (e.g., \citealt{jone25}) have used NIRSpec IFU observations of standard stars to demonstrate that the python package \textlcsc{stpsf}\footnote{\url{https://stpsf.readthedocs.io/en/latest/}} is able to accurately characterises the PSF.
Following the approach of other works (e.g., \citealt{jone25b,prie25}), we use first use \textlcsc{stpsf} to find the PSF of each spectral channel in our data cube. We then homogenise the PSF of our data cube to the coarsest PSF by convolving each spectral channel with a kernel determined using the \textlcsc{phoutils} \citep{brad23} task \textit{create\_matching\_kernel}.

Next, we determine the aperture loss correction for each aperture of Table \ref{aper_table} using the common PSF. Assuming an unresolved point source (e.g., \citealt{prie25}), we take the ratio of the total flux and the flux inside each aperture. While the higher spatial resolution JWST/NIRCam observations of \citet{suga24} found that the eastern core is indeed compact ($r_{\rm e}<0.016''$), the western core is instead elongated with a length of $\sim0.3''$ and may feature small-separation clumps. Our data feature a PSF with FWHM$\sim0.19''$, so our assumption of a point source is appropriate for the eastern core but may result in a slight underestimation of the western flux.

\section{Astrometric correction}\label{astrom_app}
In this work, we compare JWST/NIRSpec IFU data with ALMA observations. This comparison requires a trustworthy astrometric correction, as even small offsets can introduce artificial features (e.g., spatial gradients in line ratio maps). Here, we align an archival NIRCam image to the \textit{Gaia} DR3 frame (\citealt{gaia16,gaia21}) using a comparison to NIRCam data (Appendix \ref{astro1}), use this aligned NIRCam image to correct the astrometry of our NIRSpec data (Appendix \ref{astro2}), and discuss the astrometric uncertainty of the ALMA data (Appendix \ref{astro3}).

\subsection{NIRCam alignment}\label{astro1}
\begin{figure}
    \centering
    \includegraphics[width=0.5\textwidth]{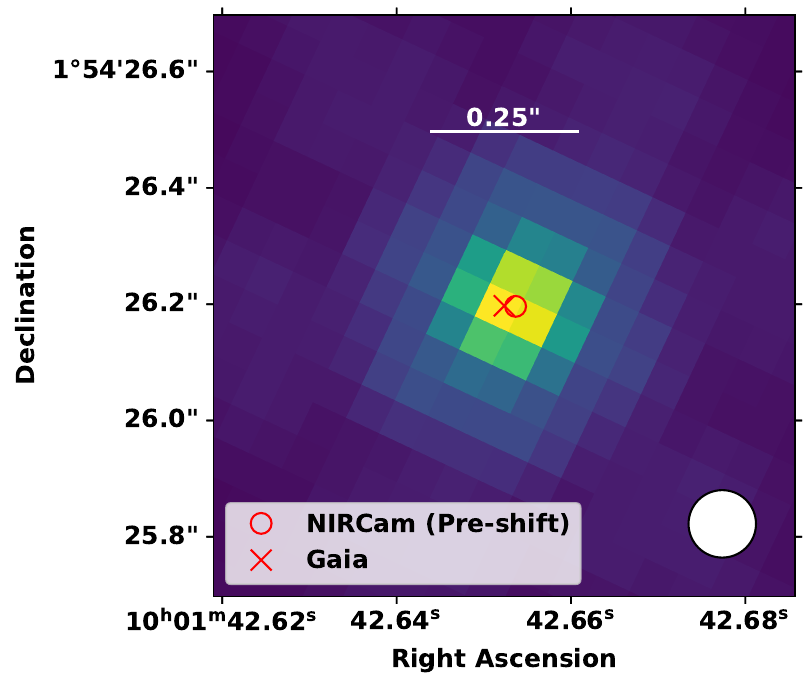}
    \caption{Comparison of JWST/NIRCam F356W data with position of star from the \textit{Gaia} DR3 catalogue. The \textit{Gaia} position is shown as a red X, while the best-fit centroid of the NIRCam image is shown by a red circle. The PSF is shown as a circle to the lower right.}
    \label{figastro}
\end{figure}

To begin, we search the \textit{Gaia} archive\footnote{\url{https://gea.esac.esa.int/archive/}} for reference stars near \bfourteen, and select the closest \textit{Gaia} star with proper motion information (Gaia DR3 3836051483531923200). Although \bfourteen benefits from multiple archival NIRCam datasets (see Section \ref{NCsec}), our chosen \textit{Gaia} star is masked in all but one filter (F356W). After correcting the position of this star for proper motion (an offset of $31.6\pm0.4$\,mas, where the uncertainty takes the positional uncertainty and error on proper motion into account) and fitting a 2D Gaussian to the NIRCam image using \textlcsc{lmfit}, we find that the centroid is offset by $20.5\pm0.7$\,mas (where the error is taken from the \textlcsc{lmfit} output). This offset is small (Figure \ref{figastro}) but significant, so we align the NIRCam data by adjusting the image header.

\subsection{NIRSpec alignment}\label{astro2}

\begin{figure}
    \centering
    \includegraphics[width=0.5\textwidth]{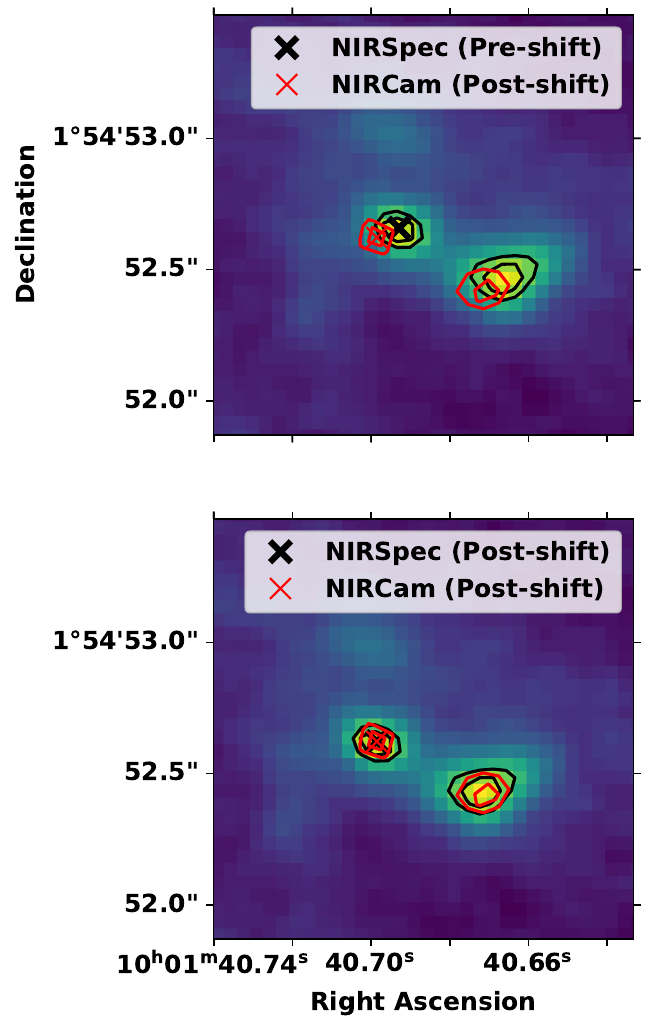}
    \caption{Comparison of JWST/NIRCam images and JWST/NIRSpec IFU data cube integrated over F356W NIRCam filter bandpass. In each panel, the collapsed NIRSpec emission is depicted as the background colour and illustrative black contours, while the NIRCam data are shown as red contours. The JWST emission in the top panel is shown without the astrometric correction, while the lower panel includes the alignment to the NIRCam data.}
    \label{astrocheck}
\end{figure}

To determine if our data are affected by a pointing error, we use the \textit{Gaia}-aligned JWST/NIRCam F356W image. By convolving our IFU data cubes with the corresponding NIRCam filter transmission curve, we compare the spatial distribution of emission. The resulting comparison is shown in the top panel of Figure \ref{astrocheck}. The NIRSpec IFU and NIRCam data are misaligned, with an offset of $92\pm8$\,mas (where the uncertainty is taken from a 2D Gaussian fit to the western emission with \textlcsc{lmfit}). This is fully consistent with the pointing uncertainty of JWST (e.g., \citealt{rigb23}), but to enable comparison with other datasets, we correct our NIRSpec IFU cube for this offset. By combining all relevant uncertainties (i.e., on \textit{Gaia} position and proper motion, NIRCam centroid, and NIRSpec centroid), we determine an absolute astrometric uncertainty of 5\,mas.

\subsection{ALMA astrometry}\label{astro3}
According to the ALMA technical handbook\footnote{\url{https://almascience.eso.org/documents-and-tools/cycle12/alma-technical-handbook}}, the positional uncertainty of ALMA is a function of the resolution and peak S/N:
\begin{equation}
\Delta_{\rm acc}=\frac{FWHM_{\rm beam}}{0.9\times SNR_{\rm pk}}
\end{equation}
Using this, we first consider the moment 0 map of \cii studied here, which features a beam of FWHM $0.268''\times0.218''$ (i.e., geometric mean of $0.242''$) and a $SNR_{\rm pk}\sim8$. This results in $\Delta_{\rm acc}=34$\,mas. Applying the same process to the moment 0 map of \oiiialma implies $\Delta_{\rm acc}=52$\,mas. However, the ALMA technical handbook notes that the true positional uncertainties are likely higher due to extended structure and high frequencies. Because the ALMA positional uncertainty may be improved through better sensitivity or resolution, future additional \cii or \oiiialma observations will result in tighter constraints.

Because these uncertainties are comparable to the spaxel size of our NIRSpec data ($50$\,mas), any detailed spatial comparison of NIRSpec and ALMA data is not advisable. However, integrated flux comparisons using large apertures are still possible. 

\section{Line ratio determination}\label{pnapp}
Throughout this work, we use line ratios derived using \pyneb. This allows us to self-consistently fit and analyse line emission using a single framework. The details of these assumptions are given here.

\subsection{Dust-free line ratios}
One vital pair of ISM properties are the electron density and temperature. The electron density in galaxies has been derived for $z\sim2.3$ galaxies ($n_{\rm e}=250$\,cm$^{-3}$; e.g. \citealt{sand16}), as well as for $z>6$ galaxies using JWST data ($n_{\rm e}=100-500$\,cm$^{-3}$; e.g. \citealt{curt23,naka23,chen24,hu24,torr24}). More generally, \citet{isob23} characterise the redshift evolution of $n_{\rm e}$, yielding a range of $n_{\rm e}\sim200-1000$\,cm$^{-3}$ for a source at $z\sim7$. Below we will explore a range of $n_{\rm e}=100-1000$\,cm$^{-3}$ (as suggested by \citealt{tang23}). 

Similarly, the electron temperature is found to be $\sim1.5\times10^4$\,K for high-redshift sources (e.g., \citealt{hsia24,hu24,torr24}). While a broad range of $T_{\rm e}=[0.5-20]\times10^4$\,K is explored by \citet{chen24}, we restrict our exploration to temperatures where \pyneb is able to calculate each dust-corrected Balmer ratio ($T_{\rm e}=[0.5-3.0]\times10^4$\,K).

Previously, \citet{suga24} assumed $T_{\rm e}=1.2\times 10^4$\,K and $n_{\rm e}=200$\,cm$^{-3}$ when interpreting JWST/NIRCam observations of \bfourteen. Through a comparison of the observed \oiiiab/\oiiialma ratio to results from CLOUDY (\citealt{ferl98,ferl17}), they were able to rule out low ($<10^2$\,cm$^{-3}$) and high electron densities ($>10^3$\,cm$^{-3}$). For our exploration, we mark the locations of these fiducial values of $n_{\rm e}=200$\,cm$^{-3}$ and $T_{\rm e}=1.2\times10^4$\,K in each plot, and the resulting ratio values are listed in Table \ref{lineapptab}.

\begin{figure*}
\centering
\includegraphics[width=\textwidth]{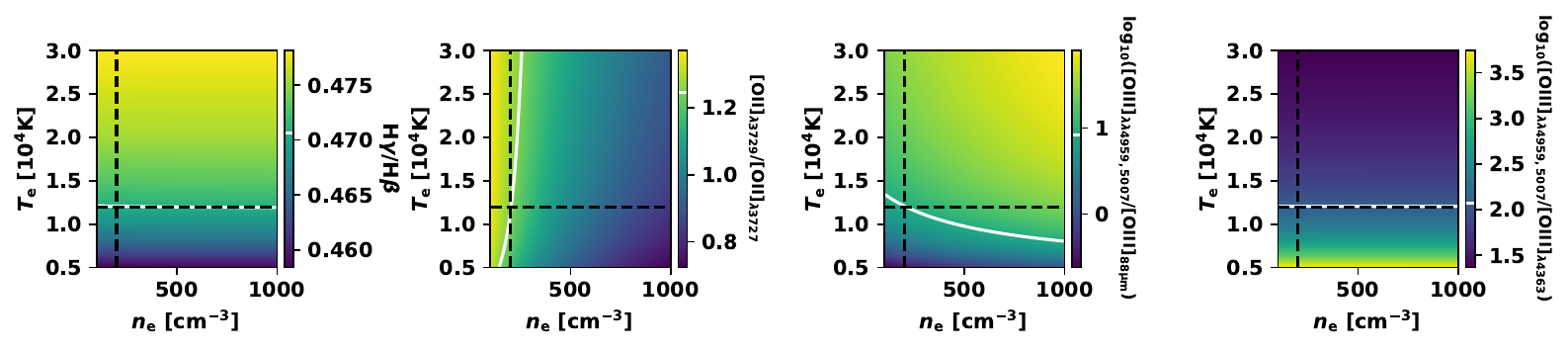}
\caption{Demonstration of how line ratios vary with respect to $T_{\rm e}$ and $n_{\rm e}$, with $E(B-V)=0.0$. The fiducial values ($n_{\rm e}=200$\,cm$^{-3}$ and $T_{\rm e}=1.2\times10^4$\,K) are marked with dashed black lines, while the resulting ratio is marked in the colourbar with a white line. White lines in each panel show ISM conditions that produce the same ratio as the fiducial values. }
\label{balmerplot}
\end{figure*}

First, we explore Balmer line ratios using the \pyneb task \textit{getEmissivity}. The resulting distribution of \hg/\hb values (first panel of Figure \ref{balmerplot}) shows that this ratio features a much stronger dependence on $T_{\rm e}$ than $n_{\rm e}$. The other Balmer ratios show a similar dependence on $T_{\rm e}$, with a $<6\%$ deviations from their fiducial value over our explored parameter space. 

Next, we explore the range of multiple oxygen line ratios. The first ratio (\oiiib/\oiiia) is fixed by atomic physics and therefore does not vary with $n_{\rm e}$ or $T_{\rm e}$. On the other hand, \oiilowb/\oiilowa is strongly dependent on $n_{\rm e}$, (\oiiib/\oiiic) is dependent on $T_{\rm e}$, and (\oiiib/\oiiialma) is dependent on both parameters. Due to the large variation in the latter two ratios (i.e., more than 1\,dex), we present the logarithm of each in Figure \ref{balmerplot}.

\begin{table}
\centering
\begin{tabular}{c|ccc}
$E(B-V)$                                                        & 0.0      & 0.1    & 0.4      \\ \hline
H$\alpha$/H$\beta$                                              & 2.828    & 3.179  & 4.517 \\
H$\gamma$/H$\beta$                                              & 0.471    & 0.448  & 0.388 \\
H$\delta$/H$\beta$                                              & 0.261    & 0.252  & 0.194 \\
H$\epsilon$/H$\beta$                                            & 0.160    & 0.147  & 0.113 \\
H$\zeta$/H$\beta$                                               & 0.106    & 0.096  & 0.072 \\
H$\eta$/H$\beta$                                                & 0.074    & 0.066  & 0.049 \\\hline
\textit{[OIII]$\lambda5007$/[OIII]$\lambda4959$}                & \textit{2.984} & \textit{2.996} & \textit{3.032} \\
\oiilowb/\oiilowa                                               & 1.246    & 1.246  & 1.247 \\
\oiiib\oiiialma                                                 & 6.282    & 4.165 & 1.213 \\
\oiiib/\oiiic                                                   & 88.623    & 93.972 & 112.035 \\ \hline
\textit{[NeIII]$\lambda3968$/[NeIII]$\lambda3869$}              & \textit{0.301} & \textit{0.305} & \textit{0.315} \\
\end{tabular}
\caption{Dust-corrected line ratios found using \pyneb using fiducial electron properties ($n_{\rm e}=200$\,cm$^{-3}$ and $T_{\rm e}=1.2\times 10^4$\,K) and either $E(B-V)$ = 0, 0.1, or 0.4. Italicised entries are independent of density and temperature.}
\label{lineapptab}
\end{table}

Like \oiiib/\oiiia, the ratio \neiiia/\neiiib is fixed by atomic physics, and thus is not dependent on $n_{\rm e}$ or $T_{\rm e}$. Because of this, we present the fiducial value in Table \ref{lineapptab} but do not show a distribution.

\subsection{Dust reddening}
\begin{figure*}
    \centering
    \includegraphics[width=\linewidth]{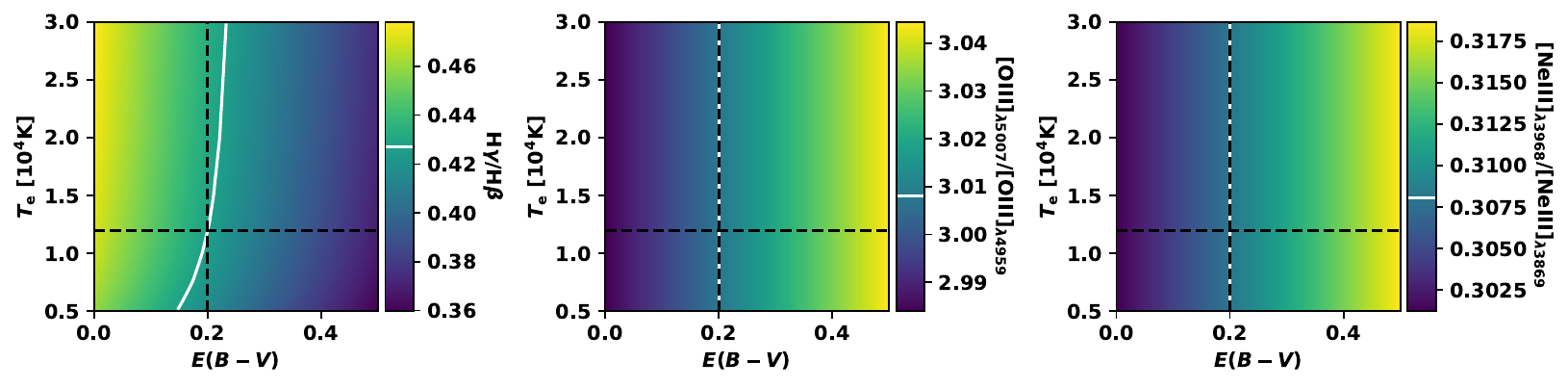}
    \caption{Demonstration of how line ratios vary with respect to $T_{\rm e}$ and $E(B-V)$, with a constant $n_{\rm e}=200$\,cm$^{-3}$. The fiducial temperature ($T_{\rm e}=1.2\times10^4$\,K) and B-V colour excess ($E(B-V)=0.2$) are marked with dashed black lines, while the resulting ratio is marked in the colourbar with a white line. White lines in each panel show ISM conditions that produce the same ratio as the fiducial values.}
    \label{tebv}
\end{figure*}

In addition, line ratios may be affected by dust extinction. In the case of a dust-free environment, $E(B-V)=0$, and the line ratios discussed in the previous subsection would be appropriate. However, even for $E(B-V)\lesssim0.5$, line ratios will differ significantly from their extinction-free values. To demonstrate this, Figure \ref{tebv} shows the $T_{\rm e}-E(B-V)$ grid for \hg/\hb, \oiiib/\oiiia, and \neiiib/\neiiia. 

For each of the Balmer ratios, over the range of parameters considered, $E(B-V)$ has more of an effect on the observed ratio than $T_{\rm e}$. This is also true for the oxygen and neon ratios, which are temperature-independent. While the oxygen and neon ratios only vary a few percent from the fiducial values ($\lesssim3\%$), the Balmer ratios have larger variations that increase with transition number (e.g., $\sim15\%$ for \hg/\hb and $\sim30\%$ for \hz/\hb). 

\subsection{Line ratio interpretation}\label{lri}
It is clear that some line ratios are strong tracers of ISM conditions. A high \oiiib/\oiiia or \neiiib/\neiiia ratio implies strong dust reddening, while in the absence of reddening a low \oiilowb/\oiilowa ratio will suggest a high electron density, and a high Balmer ratio (e.g., \hg/\hb) traces a high electron temperature. But there is some ambiguity between $n_{\rm e}$, $T_{\rm e}$, and $E(B-V)$, as shown by the curved lines of constant ratio in Figures \ref{balmerplot} and \ref{tebv}.

While \oiiib/\oiiia and \neiiib/\neiiia appear to be good tracers of $E(B-V)$, their small wavelength separations result in small changes in the observed ratio (e.g., $\lesssim1\%$ for $E(B-V)\lesssim0.5$) which would require high-S/N observations to constrain. In addition, we have assumed case B recombination here, while case A recombination would result in altered Balmer ratios (e.g., \citealt{scar24}). We fit each observed profile in this work with two components (i.e., narrow and broad), adding an additional complication to ratio interpretation. 

With these notes in mind, we make several assumptions for our fitting procedure. First, since we lack any strong tracers of $n_{\rm e}$ (i.e., the spectral resolution to resolve the \oiilow doublet), we fix $n_{\rm e}$ to the fiducial value of $200\,$cm$^{-3}$ used by \citet{suga24}. Next, we conservatively fix $E(B-V)$ to the value found by \citet{suga24} for the full \bfourteen system ($0.2$). Finally, we assume that all lines originate from the same medium, such that each line may be characterised by the same ISM conditions ($n_{\rm e}$, $T_{\rm e}$, and $E(B-V)$).

\section{Comparison of narrow and broad emission}\label{NBAPP}

\begin{figure}
    \centering
    \includegraphics[width=0.5\textwidth]{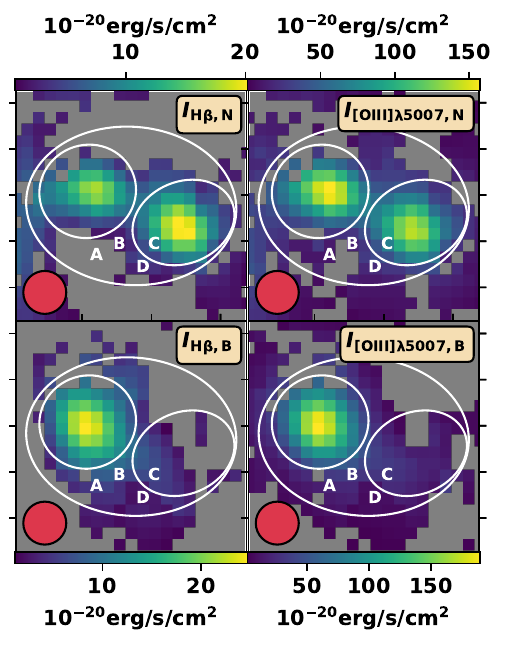}
    \caption{Integrated fluxes of narrow (top row) and broad (bottom row) components of \hb (left column) and \oiiib (right column), as derived through spaxel-by-spaxel fit. Each panel displays a field of view of $1.0''\times1.0''$ centred on 10h01m40.69s +1$^{\circ}$54$'$52.55$''$. For each, we show the PSF as a red ellipse to the lower left. North is up and east is to the left. We mark the locations of four spaxels (A through D), from which we extract spectra (Figure \ref{NB_spectral}).}
    \label{NB_spatial}
\end{figure}

In Section \ref{spaxelbyspaxel}, we extract the spectra from each spaxel of our data and fit them to determine the spatial distribution of line and continuum emission, as well as ISM properties. While most of our discussion focuses on the total emission, our model includes both a narrow and broad component. To demonstrate the significance of these detected components, we show the best-fit narrow and broad emission maps of the two brightest lines (\hb and \oiiib) in Figure \ref{NB_spatial}.

Emission from the narrow components of both lines is focused in the two cores, while the broad emission primarily emanates from the E core. We emphasise that the fitting of each spaxel-based spectrum was performed independently, and a signal-to-noise limit of $>3\sigma$ was enforced for each fit to be accepted. 

\begin{figure}
    \centering
    \includegraphics[width=0.49\textwidth]{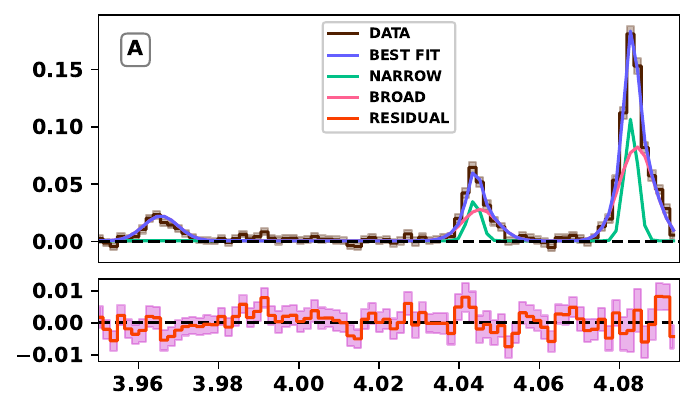}
    \includegraphics[width=0.49\textwidth]{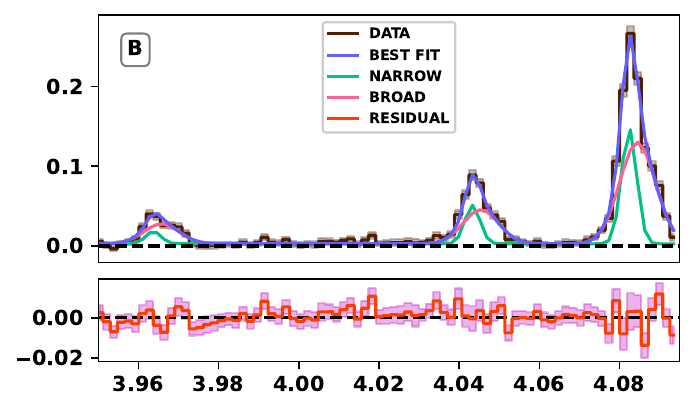}
    \includegraphics[width=0.49\textwidth]{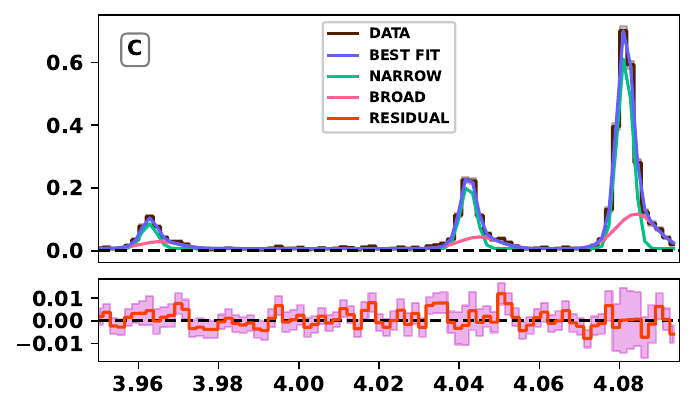}
    \includegraphics[width=0.49\textwidth]{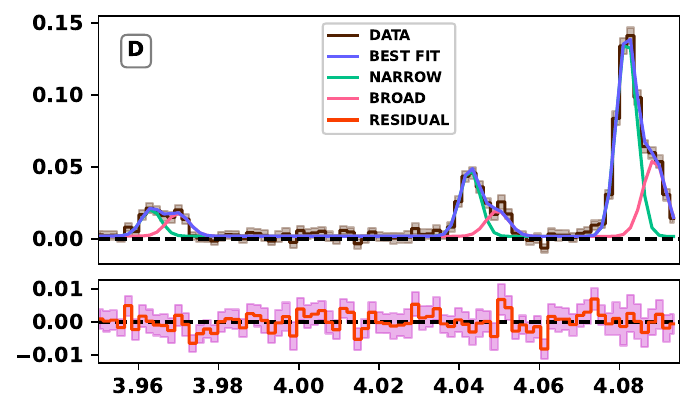}
    \caption{Spectra extracted from four spaxels of our JWST/NIRSpec IFU data cube (see Figure \ref{NB_spatial}). 
    We isolate the wavelength range containing \hb and \oiiiab, and present the observed spectrum, the best-fit model, and its narrow and broad components. The residual is included in the lower panel.}
    \label{NB_spectral}
\end{figure}

This is further demonstrated in Figure \ref{NB_spectral}, where we show extracted spectra from four spaxels between the cores (see positions A-D in Figure \ref{NB_spatial}), their best-fit models, and the residuals. Each \ha and \oiiiab line show significant broad emission, with different strengths and velocity offsets from the narrow emission.

\bsp	
\label{lastpage}
\end{document}